\documentclass[structabstract]{aa}  
\usepackage{graphicx}
\usepackage{natbib}
\usepackage{amsmath}
\usepackage{multirow}
\usepackage{lscape}
\usepackage{exscale}   
\usepackage{relsize}
\usepackage[all]{xy}  
\usepackage{url}      %

\DeclareSymbolFont{UPM}{U}{eur}{m}{n}
\SetSymbolFont{UPM}{bold}{U}{eur}{b}{n}
\DeclareMathSymbol{\umu}{0}{UPM}{"16} 

\bibpunct{(}{)}{;}{a}{}{,} 

\usepackage{txfonts}
\begin{document}

  \title{Multiple episodes of star formation in the CN15/16/17 molecular complex. \thanks{Based on observations performed at ESO's La Silla-Paranal observatory. Programme ID 085.D-0780. }}


   \author{M. Gennaro
          \inst{1}\fnmsep\thanks{Member of the "International Max Planck Research School for Astronomy 
and Cosmic Physics at the University of Heidelberg" (IMPRS-HD).}
	  \and
	  A.~Bik \inst{1}
          \and
          W.~Brandner \inst{1}
	  \and
	  A.~Stolte \inst{2}
	  \and
	  B.~Rochau \inst{1}
	  \and
	  H.~Beuther \inst{1}
	  \and
	  D.~Gouliermis \inst{1,3}
	  \and
	  J.~Tackenberg \inst{1, \star\star}
	  \and
	  N.~Kudryavtseva \inst{1, \star\star}
	  \and
	  B.~Hussmann \inst{2}
	  \and
	  F.~Schuller \inst{4}
	  \and
	  Th.~Henning \inst{1}
          }

   \institute{Max-Planck-Institut f\"ur Astronomie, K\"onigstuhl 17, D-69117, Heidelberg, Germany \\
              \email{gennaro@mpia.de}\\
         \and
	      Argelander-Institut f\"ur Astronomie, Auf dem H\"ugel 71, D-53121, Bonn, Germany\\
         \and
	      Zentrum f\"ur Astronomie der Universit\"at Heidelberg, Institut f\"ur Theoretische Astrophysik, Albert-Ueberle-Str.~2, D-69120 Heidelberg, Germany\\
	 \and	
              European Southern Observatory, Alonso de Cordova 3107, Casilla 19001, Santiago 19, Chile\\
	      }

   \date{Received 14 July 1789; accepted 20 July 1969}

 
  \abstract
   {We have started a campaign to identify massive star clusters inside bright molecular bubbles towards the Galactic Center.
   The CN15/16/17 molecular complex is the first example of our study. The region is characterized by the presence of two young clusters, DB10 and DB11, visible in the near infrared, an ultra-compact H~\textsc{ii} region identified in the radio, several young stellar objects visible in the mid infrared, a bright diffuse nebulosity at 8$\umu$m coming from PAHs and sub-mm continuum emission revealing the presence of cold dust.}
   {Given its position on the sky ($l=0\fdg58$, $b=-0\fdg85$) and its kinematic distance of $~7.5$ kpc, the region was thought to be a very massive site of star formation in proximity of the Central Molecular Zone. One of the two identified clusters, DB11, was estimated to be as massive as $\sim 10^4 \,M_{\sun}$. However the region's properties were known only through photometry and its kinematic distance was very uncertain given its location at the tangential point. We aimed at better characterizing the region and assess whether it could be a site of massive star formation located close to the Galactic Center.} 
   {We have obtained NTT/SofI deep $JHK_{\mathrm{S}}$ photometry and long slit $K$ band spectroscopy of the brightest members. We have additionally collected data in the radio, sub-mm and mid infrared, resulting in a quite different picture of the region.}
   {We have confirmed the presence of massive early B type stars and have derived a spectro-photometric distance of $\sim1.2$ kpc, much smaller than the estimated kinematic distance. Adopting this distance we obtain clusters masses of $M_{\mathrm{DB10}} \approx 170 M_{\sun}$ and $M_{\mathrm{DB11}} \approx 275 M_{\sun}$. This is consistent with the absence of any O star, confirmed by the excitation/ionization status of the nebula. No He~\textsc{i} diffuse emission is detected in our spectroscopic observations at 2.113$\umu$m, which would be expected if the region was hosting more massive stars. Radio continuum measurements are also consistent with the region hosting at most early B stars.}
   {}

   \keywords{ISM: individual: CN15/16/17 -- ISM: bubbles -- Open clusters and associations: individual: DB10, DB11 -- H~\textsc{ii} regions }
  
\maketitle 


%

\section{Introduction}

\begin{figure*}[t]
 \centering
  \resizebox{0.98\hsize}{!}{\includegraphics{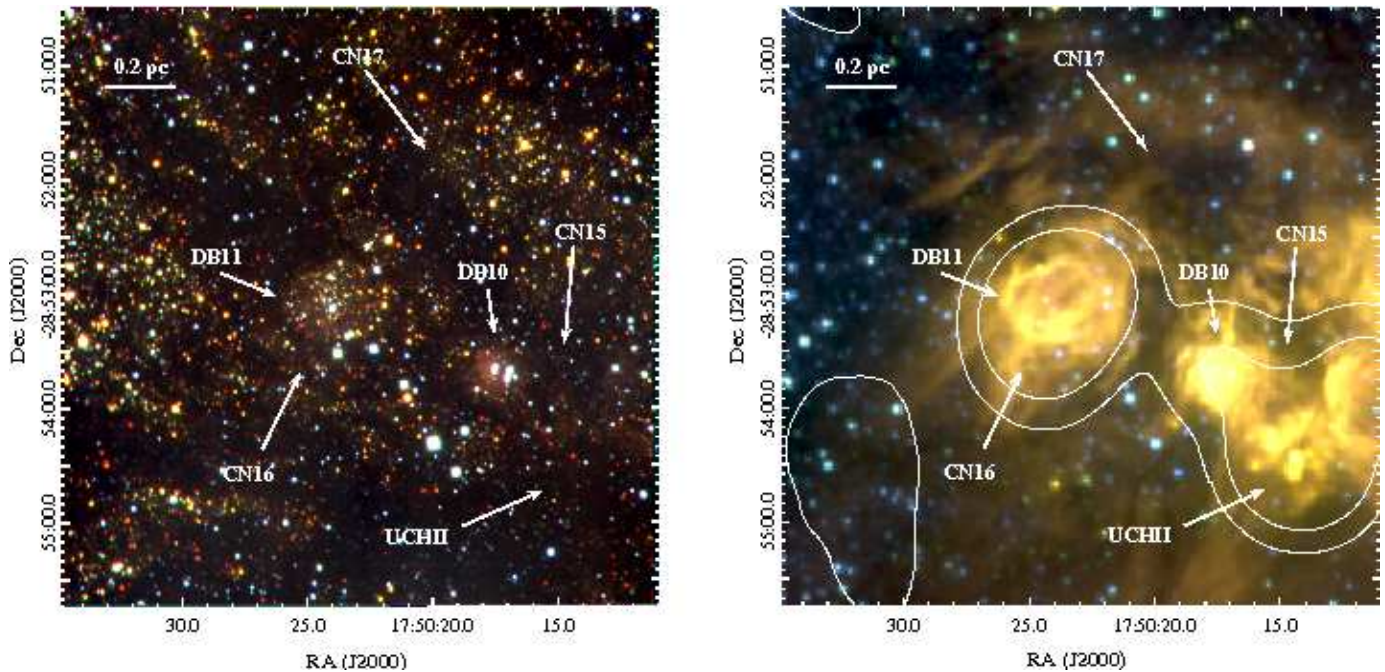}}
 \caption{Left: SofI $J$ (blue), $H$ (green) and $K_{\mathrm{S}}$ (red) composite image of the CN15/16/17 complex. Right: Spitzer/IRAC 3.6$\umu$m (blue), 5.8$\umu$m (green) and 8.0$\umu$m (red) composite image for the same region. The white contours are NVSS radio continuum maps. The physical length-scale of 0.2 pc is estimated at a distance of 1200 pc (see Sect.~\ref{sec:fsd}).}
 \label{fig:imreg}
\end{figure*}

Due to the high degree of interstellar extinction, very few young clusters are known in the Galactic Plane at distances larger than 2 kpc from the Sun. 
Nevertheless, the central region of the Galaxy, with its unique physical conditions, represents a very interesting laboratory for testing star formation theories \citep{1996ARA&A..34..645M}. Recent studies give somewhat controversial results on the star formation efficiency (SFE) in the proximity of the Galactic Center (GC): some suggest a SFE similar to the Milky Way Disk \citep{2009ApJ...702..178Y}, others imply a reduced SFE towards the GC, based on a Galactic-wide
comparison of dense gas tracers and active star formation tracers \citep{2012ApJ...746..117L}. 
These differences highlight the need for further observational constraints on the star formation scenario close to the GC.

Some studies suggest that several hundred clusters are expected to reside along the line of sight towards the inner Galaxy \citep{2001ApJ...546L.101P}, but very few of these clusters have so far been identified.
Unfortunately, if we look towards the center of the Galaxy, the identification of clusters is complicated by the severe crowding of fore/background stars.
Spurious cluster detections are also possible due to the patchy nature of the interstellar medium, which may cause strong spatially varying foreground extinction and consequently a variation in the star counts.
Due to the strong extinction along lines of sight within the Galactic Plane, the detection of distant clusters at optical wavelengths is challenging, even for the most massive clusters located towards the center of the Milky Way.
The best way to identify new clusters in the inner Galaxy is therefore to use infrared wavelengths, which are less affected by extinction.

Over the past decade, the search for new clusters has gained fresh interest thanks to near infrared surveys such as DENIS \citep{1999A&A...349..236E}, 2MASS \citep{Skrutskie:2006uq}, UKIDSS-GPS \citep{2008MNRAS.391..136L} and VISTA-VVV \citep{2007Msngr.127...28A,2010NewA...15..433M}.
In parallel to these near infrared ground based surveys, mid infrared surveys of the Galactic Plane have been carried out using the IRAC camera on board the Spitzer Space Telescope.
\cite{2006ApJ...649..759C,2007ApJ...670..428C} found almost 600 molecular bubbles in the GLIMPSE I and II surveys.
Molecular bubbles associated with H~\textsc{ii} regions are a tracer of young, massive clusters hosting O-type or early B-type stars.
Within the sample identified in the GLIMPSE II survey ($|l| < 10\degr,\, |b| < 1\degr$), 29 bubbles are associated with H~\textsc{ii} regions, meaning that they very likely host massive stars. 

We recently started a campaign aimed at characterizing the stellar content of these 29 regions, using a combination of imaging and long slit spectroscopy. In this paper we present the results for CN15/16/17. 
We observed the region using the SofI near infrared instrument mounted on the ESO-NTT telescope in La Silla, Chile. The observing strategy consisted of a combination of deep $JHK_{\mathrm{S}}$ imaging and long slit $K$ band spectroscopy of the brightest candidate members. By combining photometry and spectroscopy we have been able to confirm the presence of early B type stars and to constrain the region's distance using their spectral type classification (see Sect.~\ref{sec:fsd}).

The CN15/16/17 complex of molecular bubbles is a star forming region (SFR) hosting young stars in different evolutionary phases (see Fig.~\ref{fig:imreg}). 
The region is projected towards the Galactic Center ($l=0\fdg58$, $b=-0\fdg85$) and was first detected by \cite{2007ApJ...670..428C}, by visually searching the inner $20\degr$ of the Galaxy, using mid infrared data from the GLIMPSE~II survey \citep[][]{2003PASP..115..953B,2009PASP..121..213C}.
The Spitzer/IRAC images of the region show a very pronounced diffuse emission in the $8\umu$m channel, originating from PAH emission.
Two stellar clusters are associated with the region and are visible in the near infrared. The clusters were first identified by \cite{2000A&A...359L...9D} using 2MASS images \citep{Skrutskie:2006uq}. One of them (DB11) has already emerged from its parental cloud and therefore its stellar population is detectable in the near infrared. The second (DB10) is still deeply embedded and the high extinction only allows the detection of the brightest sources.  \cite{2003A&A...408..127D} further studied the clusters using $H$ and $K_{\mathrm{S}}$ band imaging also obtained with NTT/SofI.
In addition, a third, very deeply embedded SFR is present. This youngest region consists of a group of young stellar objects (YSOs) visible in the Spitzer images and corresponds to the IRAS 17470-2853 source. It is associated with a radio detected Ultra Compact H~\textsc{ii} region as well as several methanol masers \citep{1998MNRAS.301..640W}.

In Fig.~\ref{fig:imreg}, left panel, we show a $JHK_{\mathrm{S}}$ composite image of the region from our SofI observations. Objects DB10 and DB11 are clearly visible. DB11 is the central, larger cluster, while DB10 is the smaller cluster west of DB11. 
The Spitzer/IRAC composite image, right panel, is dominated by bright PAHs emission in the $8\umu$m channel. The contours indicate the radio continuum flux at 1.4 GHz from the NRAO VLA Sky Survey \citep[NVSS, see][]{1998AJ....115.1693C}. From the radio contours it is possible to identify two H~\textsc{ii} regions, one associated with DB11, and the other with the aforementioned Ultra Compact H~\textsc{ii} region, corresponding to the third, deeply embedded, SFR in the complex. In the latter, a group of YSOs can be seen in the IRAC channels, but most of them are too deeply embedded to be detected in the $JHK_{\mathrm{S}}$ image. In fact, only two are visible as very red sources.
Weaker, but still traceable radio emission is also observed at DB10's position, resulting in an elongation of the contours.

The paper structure is as follows: in Sect.~\ref{sec:phot} we describe reduction and photometry for the new $JHK_{\mathrm{S}}$ SofI images, Sect.~\ref{sec:IRACp} deals with YSO identification using IRAC photometry, in Sect.~\ref{sec:spec} we describe our SofI $K$ band spectroscopic observations, which are used in Sect.~\ref{sec:fsd} to derive the spectro-photometric distance of CN15/16/17, in Sect.~\ref{sec:submm} we use sub-mm data from the APEX\footnote{APEX is a collaboration between the Max-Planck-Institut f\"ur Radioastronomie, the European Southern Observatory and the Onsala Space Observatory.} Telescope Large Area Survey of the Galaxy \citep[ATLASGAL, ][]{2009A&A...504..415S} to estimate extinction towards the complex, in Sect.~\ref{sec:radio} we use integrated radio emission from NVSS to put additional constraints on the spectral type of the most massive members in the DB10 and DB11 clusters, the same is done in Sect.~\ref{sec:nebem} using the diffuse Br$\gamma$ emission from our spectra,
this collection of data is used to derive the masses of the DB10 and DB11 clusters in Sect.\ref{sec:db10db11}, we discuss the results and summarize our conclusions in Sect.~\ref{sec:conc}.

\section{\emph{JHK$_{\mathbf{S}}$} Imaging and Photometry}
\label{sec:phot}
Observations were performed on the 27th of June 2010. A summary of the integration times can be found in Table \ref{tab:obsstr}. 
The integration times (DIT) were chosen to be short enough to avoid saturation of stars brighter than 9 mag in each of the $JHK_{\mathrm{S}}$ bands.
At each position, NINT=10 frames are averaged to produce a single raw frame; NDIT=5 dithering offset positions are used. The total exposure time is DIT$\times$NDIT$\times$NINT. 

Given the SofI field of view (FoV) of $4.5\arcmin \times 4.5\arcmin$ and their projected distance on the sky of $1.6 \arcmin$, both clusters DB10 and DB11 are covered at each dither position.
We spent an equal amount of time on the target as on a nearby field located $\sim 6 \arcmin$ to the East and $\sim4 \arcmin$ to the North from the science field. This control field was observed to build an image of the sky free of the nebulous emission, which characterizes the CN15/16/17 region.
Image reduction was performed using a combination of \texttt{eclipse} routines \citep{Devillard:2001fk} and custom-made \texttt{IDL} routines. 

\begin{figure}
 \centering
  \resizebox{\hsize}{!}{\includegraphics{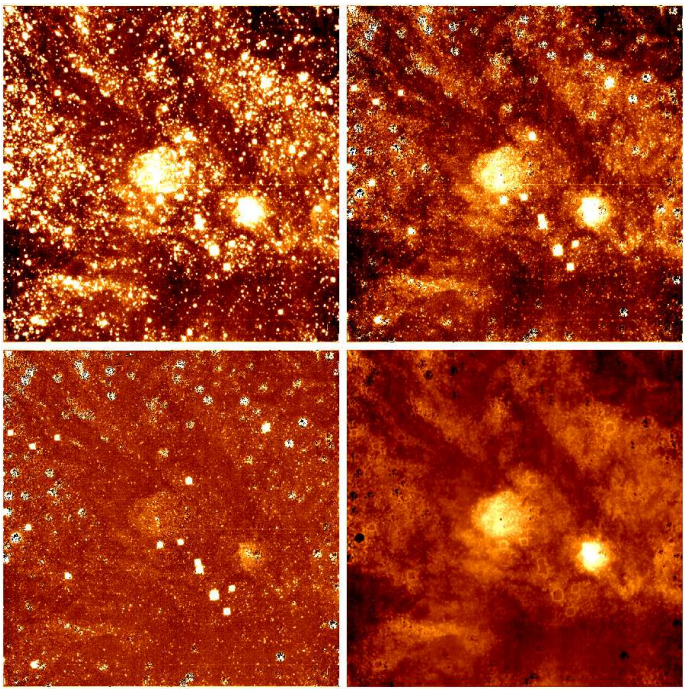}}
 \caption{Result of the image filtering. \emph{Upper left:} $K_{\mathrm{S}}$ band image. \emph{Upper right:} the same image after one \texttt{allstar} pass, with the detected sources subtracted. \emph{Lower left:} The same image after spatial filtering. \emph{Lower right:} the nebulosity that has been filtered out in the process.}
 \label{fig:FFT_res}
\end{figure}

\begin{figure*}
 \centering
  \resizebox{\hsize}{!}{\includegraphics{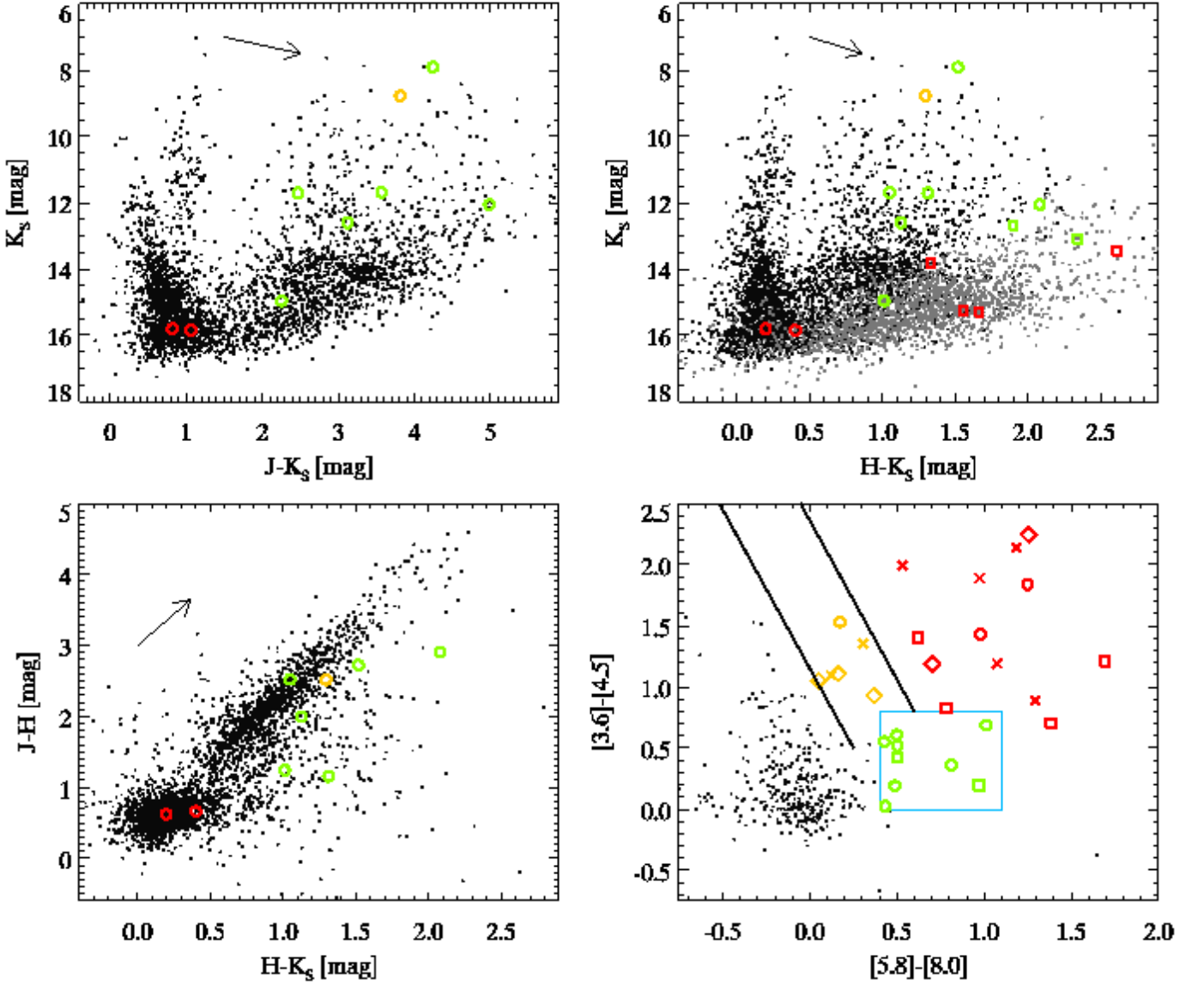}}
 \caption{Photometry of CN15/16/17. Black dots are stars identified in $J$, $H$ and $K_{\mathrm{S}}$ bands, gray dots (upper-right only) are stars identified only in $H$ and $K_{\mathrm{S}}$. YSOs are shown in red (class I), green (class II) and yellow (reddened class II). Among the YSOs, the circles are sources with $JHK_{\mathrm{S}}$ photometry, the squares are objects with only $HK_{\mathrm{S}}$ detections, the diamonds are objects with only $K_{\mathrm{S}}$ and the crosses are objects detected neither in $J$, $H$ or $K_{\mathrm{S}}$.
 The black arrows in the near infrared diagrams are the reddening vectors for $A_{K_{\mathrm{S}}} = 0.5$ mag, according to the \cite{1985ApJ...288..618R} reddening law (suitable for these wavelengths).
 \emph{Upper left:} $K_{\mathrm{S}}$ vs. $J-K_{\mathrm{S}}$ diagram. \emph{Upper right:}  $K_{\mathrm{S}}$ vs. $H-K_{\mathrm{S}}$ diagram.  \emph{Lower left:} $J-H$ vs. $H-K_{\mathrm{S}}$ diagram. \emph{Lower right:} IRAC [3.6]-[4.5] vs [5.8]-[8.0] color-color diagram. The cyan box indicates the locus of class II objects according to \protect\cite{2004ApJS..154..363A}. The two black lines are parallel to the reddening vector derived by the \protect\cite{1990ARA&A..28...37M} extinction law (suited for mid infrared wavelengths) and enclose the region of reddened class II objects according to \protect\cite{2004ApJS..154..367M}.}
 \label{fig:Diagram_ALL}
\end{figure*}

Photometry was performed on the reduced frames using the \texttt{IRAF} implementation of the \texttt{DAOPHOT} package \citep{Stetson:1987qy}. Due to the bright and highly variable background, especially in the $K_{\mathrm{S}}$ band, the detection of faint sources in some parts of the image was problematic. In order to tackle the problem, we performed spatial filtering of the image using the Fast-Fourier-Transform \citep[FFT,][]{FFTart} method in \texttt{IDL}.
After running \texttt{allstar} on the frames once, the task \texttt{substar} automatically removes the detected sources producing an image which consists of three components: the light from the missed sources $\mathcal{S}(x,y)$, the nebular diffuse emission $\mathcal{N}(x,y)$ and the residual noise from the bright infrared sky background $\mathcal{B}(x,y)$. 

We used the following procedure to reduce the $\mathcal{N}(x,y)$ term. 
In the previous steps we built a model for the stellar point spread function (PSF) using bright, isolated sources in the field. The FFT of the PSF has been used as a high-pass filter to remove the power associated with small wave numbers (i.e. the large-scale variations of the diffuse nebulosity).
Spatial filtering consists of multiplying the FFT of the image with the FFT of the PSF, and performing an inverse FFT on this product. In the final image the $\mathcal{N}(x,y)$ component is significantly reduced. The high-pass filtering removes most of the power associated with wave numbers $k \lesssim 1/\Lambda$, where $\Lambda$ is the typical length scale of the PSF (for example the full width at half-maximum (FWHM) in the case of a Gaussian PSF).

The spatial filtering does not account for the $\mathcal{B}(x,y)$ noise and the Poisson noise associated with the $\mathcal{N}(x,y)$ signal. Both of these components are indeed highly variable (i.e. their characteristic length-scale is $l \lesssim \Lambda$).
The presence of these two components sets the fundamental detection limit for the faint sources.
After applying spatial filtering a second run of the \texttt{daofind} peak-finding routine allows detection of additional sources, without the need of artificially lowering the detection threshold. Only the peak-finding step is run on the processed images. The PSF-fitting task \texttt{allstar} is subsequently performed on the original, unsubtracted and unfiltered image, using an input list obtained by merging the previous detections and the newly found peaks.
We iterate this procedure one more time to obtain the final list of instrumental stellar magnitudes in each band. 

Figure \ref{fig:FFT_res} shows the images before and after the spatial filtering is applied. The upper-left panel shows the original image in the $K_{\mathrm{S}}$ band. The upper-right panel is the result after one PSF-fitting iteration with \texttt{allstar}. The detected sources are fitted and subtracted from the original image. The residual light is coming from the undetected sources, the nebula and the sky background. The lower-left panel displays the subtracted image after spatial filtering and the lower-right panel shows the subtracted nebulosity. Schematically we have that:
\[
\xymatrix{
        \mbox{Upper Left} \ar[r]^{\quad \mathtt{allstar} \;\quad}  & \mbox{Upper right}\ar[r]^{\; \mathrm{FFT} \; } & \mbox{Lower left}\ar[d] \\
        \quad \quad \phantom{0} \ar[u]     &\ar[l] \quad \mbox{Peak finding} \quad &\ar[l]\quad\quad \phantom{0} \quad }
\]

Note that the image cut levels and color-scaling parameters in \ref{fig:FFT_res} are chosen to be the same in all panels to demonstrate the brightness of the subtracted nebulosity and the dimension of its luminosity contrast between different areas.

As already mentioned, our spatial filtering method is not removing variations in $\mathcal{N}(x,y)$ on spatial scales of the order of $\Lambda$.
Therefore some of the detected sources in a single band might be of non stellar origin (i.e. small-scale, large-gradient variations in the diffuse nebulosity) which are removed from the final catalog by matching the three separate lists of detections for $J$, $H$ and $K_{\mathrm{S}}$.
For the positional matching we use the $K_{\mathrm{S}}$ band detection list as a master catalogue and assume that any real object detected in the $H$ band must have a $K_{\mathrm{S}}$ band counterpart and any J band detection must have a counterpart in both H and $K_{\mathrm{S}}$ bands.
A total of 3332 stars are detected in $J$, $H$ and $K_{\mathrm{S}}$ bands and additional 1823 stars are detected in  $H$ and $K_{\mathrm{S}}$ only.

Instrumental magnitudes have been transformed into the 2MASS photometric system by matching our detections with bright and isolated 2MASS sources. 
A linear fit of the 2MASS vs. instrumental magnitudes was performed between the matched stars, allowing for a color term.
The resulting photometry for the entire region is displayed in Fig.~\ref{fig:Diagram_ALL}.

\begin{table}
 \caption{Observing log for the $JHK_{\mathrm{S}}$ observations.}
 \centering
 \begin{center}
\begin{tabular}{lcc}
\hline
\hline
 & \\[-2mm]
Band              & Exposure Time  & Seeing \\
		  &  [s]    & [arcsec]  \\
 & \\[-2mm]
\hline
 & \\[-2mm]
$J$               & 500 & 1.2\\
$H$               & 300 & 1.1\\
$K_{\mathrm{S}}$  & 125 & 1.0\\
 & \\[-2mm]
\hline
 \end{tabular}
 \end{center}
\label{tab:obsstr}
\end{table}


\section{IRAC mid infrared photometry}
\label{sec:IRACp}

In addition to the near infrared photometry from SofI, we analysed the mid infrared photometry for the region in the 4 IRAC bands centered at 3.6 $\umu$m, 4.5$\umu$m, 5.8$\umu$m and 8.0$\umu$m. The region has been observed within the GLIMPSE II survey \citep{2003PASP..115..953B,2009PASP..121..213C}. In this work, we used the GLIMPSE II Epoch 2 November '09 Point Source Catalog\footnote{Available at the following URLs: \\ irsa.ipac.caltech.edu/data/SPITZER/GLIMPSE/ (images) \\ irsa.ipac.caltech.edu/cgi-bin/Gator/nph-scan?submit=Select\&projshort=SPITZER (catalogs)}.

In an area of $6\arcmin \times 6 \arcmin$, centered on DB11, 342 sources are detected in all of the 4 IRAC bands. Among these objects, class I and class II YSOs can be classified using their mid infrared excess due to the circumstellar envelopes and disks, respectively. \cite{2004ApJS..154..363A} showed that these two classes of objects occupy different areas in the [3.6]-[4.5] vs. [5.8]-[8.0] color-color diagram with respect to ''purely photospheric'' objects which are clustered around the (0,0) position. Even though there might be a small overlap between the class I and class II objects, \cite{2004ApJS..154..367M} proposed a working criterion to identify the two classes of objects and separate them from the more evolved, envelope and disk-free stellar sources. Using the models of the emission of class I and II objects by \cite{2004ApJS..154..363A}, they identify a rectangular area with $0.0 \,\mbox{mag}\, \le [3.6]-[4.5] \le 0.8\,\mbox{mag}$ and $0.4 \,\mbox{mag}\,\le [5.8]-[8.0]\le 1.1\,\mbox{mag}$ as the locus of class II YSOs. This box is displayed in the lower-right panel of Fig.~\ref{fig:Diagram_ALL}. The area of the diagram above and to the right of the box is occupied by class I objects. Part of the class II objects resides in a reddening stripe on the upper left of the unreddened class II box. The two parallel lines in the same figure show the direction of the reddening vector for the \cite{1990ARA&A..28...37M} extinction law.

We identify  13 class I, 9 class II and 6 reddened class II objects in our sample. 
\emph{Class I}: 2 sources have $JHK_{\mathrm{S}}$ photometry, 4 $HK_{\mathrm{S}}$ photometry and a further two are only detected in $K_{\mathrm{S}}$ band. 
\emph{Class II}: of the 9 class II sources 7 have $JHK_{\mathrm{S}}$ photometry and 2 have $HK_{\mathrm{S}}$ photometry. 
\emph{Reddened class II}: these object have very large visual extinction values. A reddening of $~0.4$ mag in [3.6]-[4.5] corresponds to about 30 mag in $A_{V}$. 
The average color of these reddened sources is [3.6]-[4.5] $\sim1.1$ mag, while for the unreddened class II objects an average [3.6]-[4.5] $\sim0.4$ mag is observed. This typical $\sim 0.7$ mag reddening corresponds to  $A_J \approx 14.8$ mag, $A_H \approx 9.2$ mag and $A_{K_{\mathrm{S}}} \approx 5.9$ mag, using the \cite{1985ApJ...288..618R} extinction law. Such large extinction values hamper the detection of reddened class II objects in our SofI observations: one of them is detected in $JHK_{\mathrm{S}}$, while 3 are detected in $K_{\mathrm{S}}$ band only. 

Given that the region is projected close to the Galactic Center ($l=0\fdg58$, $b=-0\fdg85$), some contamination from asymptotic giant branch (AGB) stars might be present in the photometric YSO sample. According to  \cite{2008AIPC.1001..331M}, the loci of oxygen rich AGB stars and supergiants slightly overlap with class II YSOs. Only spectroscopy of these sources could help discriminate between different stellar types. The class I sample, on the other hand, should not be contaminated.

\begin{figure}
 \centering
  \resizebox{\hsize}{!}{\includegraphics{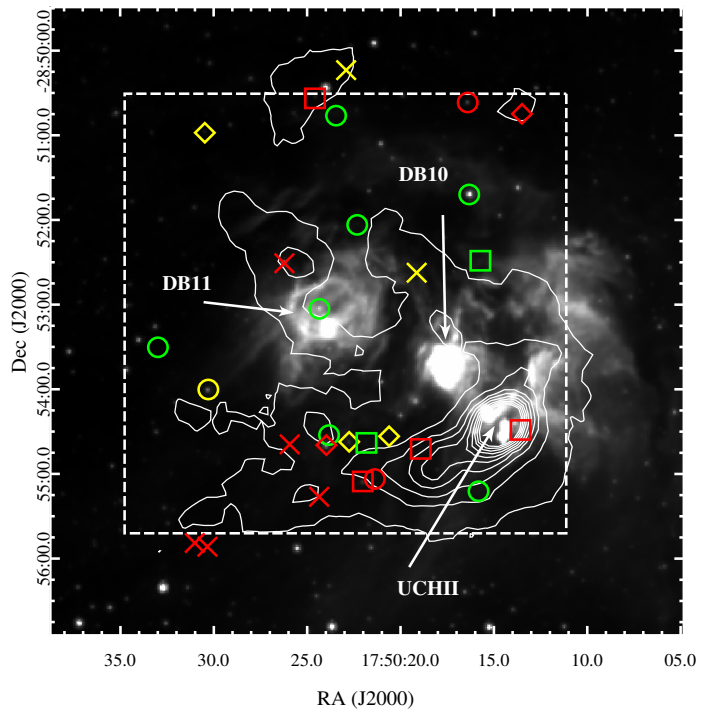}}
 \caption{IRAC 8 $\umu$m image with the marked positions of the identified YSOs. Crosses correspond to objects which are not identified in the $JHK_{\mathrm{S}}$ bands. Diamonds are objects identified only in $K_{\mathrm{S}}$, squares are objects identified in $H$ and $K_{\mathrm{S}}$, circles are objects identified in $J$, $H$ and $K_{\mathrm{S}}$. Class I objects are in red, class II in green, reddened class II sources in yellow. The dashed box corresponds to the observed SofI field. The contours trace the emission of cold dust at 870 $\umu$m from the ATLASGAL survey of the Galactic Plane. Contour levels are between 0.5 and 9.5 Jy/beam in steps of 1.0 Jy/beam.}
 \label{fig:YSOreg}
\end{figure}

The IRAC 8$\umu$m image of the region is presented in Fig.~\ref{fig:YSOreg} with the positions of the identified YSOs superimposed. Their spatial distribution shows that star formation is occurring throughout the entire region and not only in the DB10 and DB11 clusters. For example, a small clustering of YSOs is visible towards the southern part of the region. This area appears very dark in the near infrared images, due to extinction from the parental molecular cloud (see Sect. \ref{sec:submm}).
The DB11 and DB10 clusters show a paucity of YSOs. Part of the reason for this could be that the two clusters are evolved and effectively not hosting any YSO. 
However, the lack of YSOs might be partially due to the high brightness of the nebula in the mid infrared at the positions of DB10 and DB11, which hampers source detection.

\section{\emph{K} band spectroscopy}
\label{sec:spec}
Spectroscopic observations with SofI were performed on the 27th of June 2010. The typical seeing during the observations varied between $1\farcs2$ and $1\farcs6$. The spectroscopic targets were previously selected from the 2MASS point source catalog by choosing the brightest stars at the DB10 and DB11 cluster locations. We aimed to observe massive stars in the CN15/16/17 region and, by classifying their spectral type, infer the region's distance as well as the extinction towards those objects. To minimize the contamination by foreground main sequence stars, only candidates with $H-K_{\mathrm{S}} > 0.3$ mag were considered. However, it was still possible for field giants to contaminate the sample, as they have similar photometric properties as more distant and more reddened massive main sequence stars.
Five different slit positions were used; the positions were chosen in order to observe the 10 brightest stars in the region, but given the high degree of crowding more stars fell into the slits and we could extract spectra for 3 additional stars with sufficient signal-to-noise ratio (SNR) to allow for spectral classification.

We used the spectroscopic mode of NTT/SofI selecting the third order of the HR Grism, with a dispersion of 4.63 \AA/pixel in the 2.00-2.30$\umu$m wavelength range. We always used an $0\farcs6$ slit, resulting in a spectral resolution of 2200.
Several lines that allow for spectral classification of early type stars were present in the chosen spectral range, such as Br$\gamma$ (H~\textsc{i}, 2.166$\umu$m), and He~\textsc{i} lines (2.059, 2.113$\umu$m). Other lines which were present allowed for the classification of the late type giants contaminants, such as Na~\textsc{i} (2.206, 2.209$\umu$m) and Ca~\textsc{i} (2.261, 2.263, 2.265$\umu$m), as well as the CO absorption band at 2.3$\umu$m.

\begin{figure}
 \centering
  \resizebox{\hsize}{!}{\includegraphics{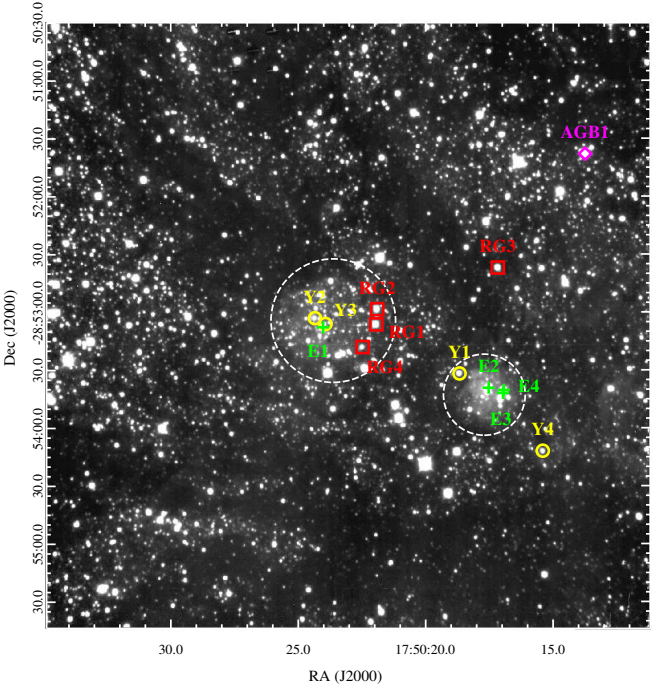}}
 \caption{$K_{\mathrm{S}}$ band image of CN15/16/17. Markers show the positions of the stars with spectral classifications. Green crosses are early type stars, yellow circles are candidate YSOs, red squares are red giants, and the magenta diamond is an AGB star. IDs correspond to the first column of Table \ref{tab:spclass}. The two white dashed circular regions are the areas corresponding to clusters DB10 (lower right, $21\arcsec$ in radius, i.e. 0.12 pc at 1200 pc distance) and DB11 (center, $32\arcsec$ in radius, i.e. 0.18 pc at 1200 pc distance).}
 \label{fig:Kb_sp}
\end{figure}

For each slit multiple nodding cycles (NC) were observed. The slit was first put in position A and then shifted along its long axis to position B. Removal of the sky background was performed using the difference between the A and B positions.
The number of NC was optimized, depending on the brightness of the sources, to obtain a SNR sufficient to perform spectral classification. 

Reduction was performed using the ESO \texttt{gasgano} pipeline \citep{2004Msngr.117...33I}. The pipeline applied flat fielding and dark subtraction for each raw image, and used xenon and neon arc lamp images to correct for slit curvature. 
For each nodding cycle the (A-B) + (B-A) sky-subtracted images were obtained. The final step was to shift and sum of the $2\times\mathrm{NC}$ sky-subtracted images.  
Wavelength calibration was also done using the xenon and neon arc lamp spectra.
To extract the spectra we used the \texttt{IDL} - \texttt{optspecextr} optimal extraction package\footnote{The software is publicly available at: \\
http://physics.ucf.edu/~jh/ast/software/optspecextr-0.3.1/doc/}. This software traces the spectra and extracts the signal using the algorithm of \cite{1986PASP...98..609H} to optimize the SNR ratio of the final spectrum.

Throughout the observing run we obtained spectra of late B and early A type reference stars in order to correct for telluric absorption. These stars have the advantage of an almost featureless spectrum in the $K$ band. The positions of the standard stars were chosen to minimize the airmass difference with the CN15/16/17 targets. Telluric standards were observed about every 30 minutes in order to have similar atmospheric conditions between the standards and the targets. The only prominent feature in these early type telluric standards is their Br$\gamma$ photospheric absorption. Therefore the extracted standard star spectra were first corrected by removing the Br$\gamma$ line using the \texttt{IRAF/splot} task to fit a Voigt profile to the Br$\gamma$ line and subtract the result of the fit from the spectrum.
We then used the \texttt{IRAF/telluric} task to optimize the telluric correction. The spectrum of each target star was divided by the telluric standard star spectrum (with Br$\gamma$ subtracted). The \texttt{telluric} task allows iterative optimization of the atmospheric correction to compensate for slight variations in airmass and atmospheric conditions between the science target and the standard star observation. This is accomplished by scaling the amount of atmospheric absorption and shifting the spectrum in wavelength until a satisfactory correction is achieved.

The SNR per spectral resolution element required to achieve a good spectral classification was of the order of 100. Due to bad seeing conditions throughout the observing run, the achieved SNR is lower, ranging between 50 and 80 (depending on the stars). This led to a typical uncertainty of $\sim2$ spectral subclasses for the classification of the early type stars in our sample.

\section{Spectral Classification: distance and extinction to CN15/16/17}
\label{sec:fsd}
\begin{figure}
 \centering
  \resizebox{0.9\hsize}{!}{\includegraphics{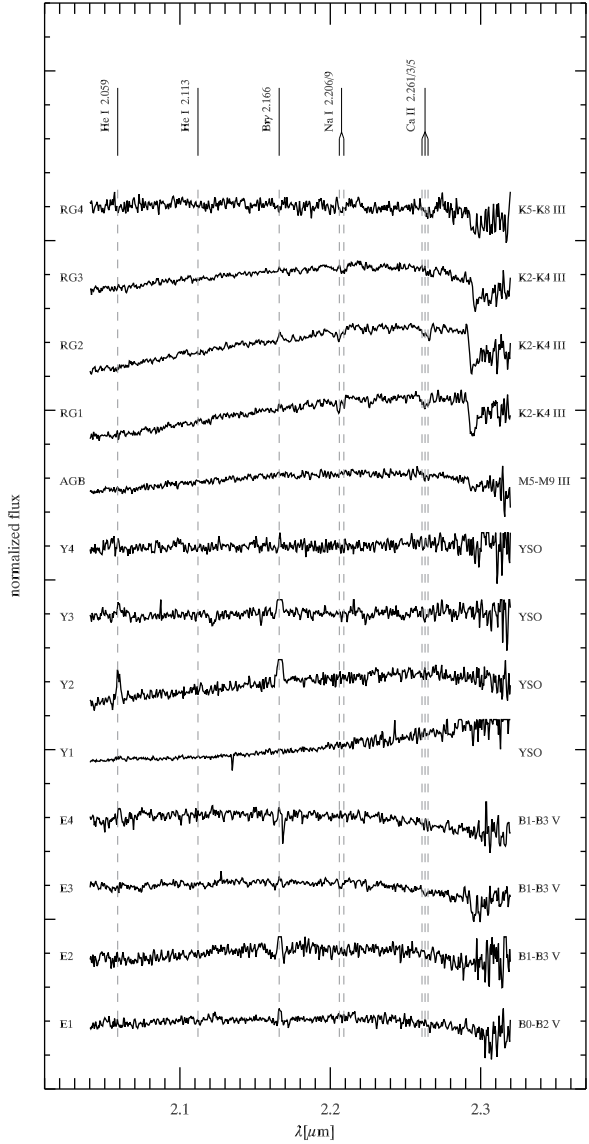}}
 \caption{Spectra of the classified objects. The IDs correspond to the first column of Table \ref{tab:spclass}. }
 \label{fig:sp_cl}
\end{figure}

The stars for which spectral classification was possible are marked in Fig.~\ref{fig:Kb_sp}, superimposed on the SofI $K_{\mathrm{S}}$ band image of the region.
The extracted and corrected spectra are shown in Fig.~\ref{fig:sp_cl}.
A total of four early B-type stars were identified and four additional YSOs candidates. The rest of the classified stars are background giants, four classified as red giants and one as an AGB star.
The results of the classification procedure are summarized in Table \ref{tab:spclass}.

Classification of the early type stars was done by comparing the observed spectra with the medium resolution near infrared spectral atlas of O and early B Stars by \cite{2005ApJS..161..154H}, complemented by spectra from \cite{2005A&A...440..121B}.
To compare these high quality spectra with our lower SNR, lower resolution ones we used the following procedure: we first convolved the atlas spectra with a Gaussian kernel with a $\sigma$ equal to 4.6 \AA/pixel, i.e. the SofI wavelength dispersion in $K$ band. 
Second we measured the SNR in a featureless part of the observed spectrum and added noise to the atlas spectra until they reached a comparable SNR.

\begin{figure}
 \centering
  \resizebox{\hsize}{!}{\includegraphics{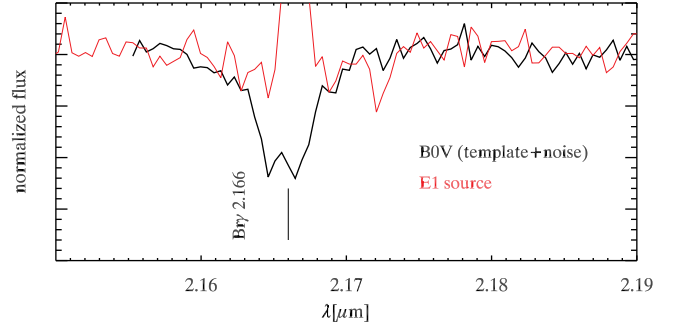}}
 \caption{The Br$\gamma$ line spectral region for the E1 source compared to the degraded spectrum of a B0V star template.} 
 \label{fig:spat}
\end{figure}

The spectral classification was then performed by visual comparison of the observed spectra with these degraded atlas spectra.
The strength of the Br$\gamma$ absorption line and of several helium lines in the observed spectral range (He~\textsc{i} 2.059$\umu$m, 2.113$\umu$m) was used to get the best match. The Br$\gamma$ absorption lines of the early type stars inside DB10 and DB11 are contaminated by the nebular Br$\gamma$ emission from hydrogen recombination. This emission, though, is quite narrow and does not affect the wings of the photospheric absorption Br$\gamma$ lines. Hence the absorption wings can still be used for spectral classification \citep[see][for a similar classification procedure]{2012ApJ...744...87B}.
An example of such comparison of the Br$\gamma$ absorption line wings is shown in Fig.~\ref{fig:spat} where we show the Br$\gamma$ line for the E1 source, classified as a B0V-B2V star, and the degraded atlas spectrum for a B0V type star.

Given the spectral types derived for the early type stars, it is possible to obtain an estimate of their distances and individual reddening values.
Using the intrinsic colors of early type stars by \cite{1995ApJS..101..117K} we determined the color excess of the observed stars. Adopting the \cite{1985ApJ...288..618R} extinction law we then estimate the foreground extinction. Given the low values of the color excess (see Table \ref{tab:spclass}), the choice of the reddening law was not critical, and the differences expected from different extinction laws are small compared to the other sources of uncertainty.
The distance was derived by the difference between the dereddened magnitude and the absolute magnitude expected for the given spectral type.
For the latter we used \cite{2001A&A...366..538L} models. We checked that the results did not depend on the actual value of the age of the isochrone adopted, in the range 1 to 3 Myr. Results in Table \ref{tab:spclass} are for the \cite{2001A&A...366..538L} ZAMS of solar metallicity. The error in the estimated extinction and distance mainly comes from the uncertainty in the derived spectral type.


We estimate an average distance to the region using the spectro-photometric values from our spectral classification. We used the reasonable assumption that the four stars are all located at the same distance from the Sun. By considering the central value of the distance interval for each of the four classified early B stars we obtained $d = 1.2 \pm 0.5 $kpc. The quoted uncertainty is the standard deviation of the four central values.  
The total visual extinction for the only B star identified in DB11 (E1) is $A_V =4.2$ mag, while the three B stars identified in DB10 (E2, E3, E4) provide a slightly larger average value of the extinction of $A_V \sim5.5$ mag, consistent with DB10 being younger and more embedded.

The identified YSOs are characterized by featureless, red spectra (see Fig.~\ref{fig:sp_cl}). Y2, Y3, Y4 show Br$\gamma$ emission and Y2 also shows He I emission at 2.059 and 2.113$\umu$m. This emission might come from the nebula, but might also come from the circumstellar material, with a similar mechanism as observed in Be stars \citep{2000A&AS..141...65C}.

For the classification of giants we used the atlas by \cite{2009ApJS..185..289R}. Given the similar resolution between our observations (R = 2200) and the spectra of this atlas (R = 2000), we could classify them by direct comparison of the observable features, such as the CO absorption band at 2.3$\umu$m, the Na~\textsc{i} (2.206, 2.209$\umu$m) and the Ca~\textsc{i} (2.261, 2.263, 2.265$\umu$m) absorption lines.

\begin{table*}\footnotesize
 \caption{Spectral Types of the classified stars. For early type stars the inferred extinction and distances are indicated as well.}
 \centering
 \begin{center}
\begin{tabular}{ccccccccccc}
\hline
\hline
& \\[-2mm]
\multirow{2}{*}{ID} & R.A.(2000) & Dec.(2000) & \multirow{2}{*}{Spectral Type} & $K_{\mathrm{S}}$ & $J-K_{\mathrm{S}}$  & $H-K_{\mathrm{S}}$ &$A_V$ &$A_{K_{\mathrm{S}}}$ & Distance \\
& $^{\mathrm{h}} \;^{\mathrm{m}}\;^{\mathrm{s}} $& $\degr\; \arcmin \;\arcsec$ &  &[mag] & [mag] & [mag] & [mag] & [mag] & [kpc] \\
& \\[-2mm]
\hline
& & & \\[-2mm]
\multicolumn{6}{l}{ Early Type } \\
& & & \\[-2mm]
E1 & 17:50:24.03 & -28:53:07.8  & B0V -- B2V &      10.36 &      0.73 & 0.21 & $4.2 \pm 0.1$  & $0.45\pm0.01$ & 1.3 -- 2.4 \\
E2 & 17:50:17.53 & -28:53:39.2  &B1V -- B3V &      10.44 &      0.87 & 0.32 & $5.7 \pm 0.1$ & $0.62\pm0.02$ & 1.0 -- 1.6 \\
E3 & 17:50:16.94 & -28:53:40.4  & B1V -- B3V &       9.75 &      0.93 & 0.26 & $4.7 \pm 0.1$  & $0.51\pm0.01$ & 0.7 -- 1.2 \\
E4 & 17:50:16.97 & -28:53:41.9  & B5V -- B9V &      11.08 &      0.89 & 0.36 & $6.0 \pm 0.1$  & $0.65\pm0.02$ & 0.6 -- 1.3 \\
& & & \\[-2mm]
\multicolumn{6}{l}{ Young Stellar Objects } \\
& & & \\[-2mm]
Y1 & 17:50:18.68 & -28:53:31.8  &red spectrum & 10.82 & 4.63 & 1.51 &-- & -- & -- \\ 
\multirow{2}{*}{Y2} & \multirow{2}{*}{17:50:24.35} &\multirow{2}{*}{-28:53:03.1} & Br$\gamma$, He~\textsc{i} 2.059, 2.113 $\umu$m  &  \multirow{2}{*}{11.70} &  \multirow{2}{*}{2.46} &  \multirow{2}{*}{1.31} &\multirow{2}{*}{--} & \multirow{2}{*}{--} \\ 
  & & & (all in emission)  & & \\
Y3 & 17:50:23.94 & -28:53:06.1 &featureless, Br$\gamma$ emission & 13.49 & 1.67 & 0.96  & --& -- & -- \\ 
Y4 & 17:50:15.41 & -28:54:11.9 &featureless, Br$\gamma$ emission & 11.48 & 0.97 & 0.36  & --& -- & -- \\ 
& & & \\[-2mm]
\multicolumn{6}{l}{ Asymptotic giant branch }   \\
& & & \\[-2mm]
AGB1 & 17:50:13.75& -28:51:38.0 & M5 -- M9 III & 8.86 & 5.35 & 1.80 &-- & --& -- \\
& & & \\[-2mm]
\multicolumn{6}{l}{ Red Giant Branch }   \\
& & & \\[-2mm]
RG1 &  17:50:21.94 & -28:53:06.3 &K2 -- K4 III &9.73  &  4.56 & 1.49 &-- & --& -- \\
RG2 &  17:50:21.90 & -28:52:58.5 &K2 -- K4 III &9.99  &  4.82 & 1.78 & -- & --& -- \\
RG3 &  17:50:17.17 & -28:52:37.0 &K2 -- K4 III &9.98  &  4.73 & 1.55 &-- & --& -- \\
RG4 &  17:50:22.48 & -28:53:18.0 &K5 -- K8 III &11.55 &  3.53 & 1.18 &-- & --& -- \\
& \\[-2mm]
\hline
\end{tabular}
 \end{center}
\label{tab:spclass}
\end{table*}

\section{Sub-mm continuum emission}
\label{sec:submm}

The cold dust emission at 870$\umu$m from the ATLASGAL survey of the Galactic Plane \citep{2009A&A...504..415S} can be used to trace extinction. The 870$\umu$m dust emission from the region is overplotted in Fig.~\ref{fig:YSOreg}.
The cold dust traces the positions of the identified YSOs in the southern part of the CN15/16/17 complex very well. Also the more isolated YSOs in the northern area are surrounded by emission at 870$\umu$m.
The peak of 870$\umu$m emission is coincident with the bright Ultra Compact H~\textsc{ii} region in the south-west of the field. This dense material is therefore associated to a region of high mass star formation. The remaining YSO population in the south of the field follows the arc-like southern region of 870$\umu$m emission and is not associated with H~\textsc{ii} emission. This is visible in Fig.~\ref{fig:imreg}, where no radio emission can be observed in this area. Hence these YSOs are likely less massive objects, and are not able to ionize their surroundings.

Using equation (1) of \cite{2009A&A...504..415S} with the derived spectro-photometric distance of 1.2 kpc and assuming $\kappa_\nu = 1.85 \mathrm{cm}^2\, \mathrm{g}^{-1}$, we estimate a mass of $\approx 105 M_{\sun}$ for the south-west clump associated with the Ultra Compact H~\textsc{ii} region. The mass of the whole arc is estimated to be $\approx 230 M_{\sun}$ by integrating the flux over the region for which the intensity is larger than 2 Jy/beam. Given the presence of nearby hot stars, we assumed a dust temperature of 50K, keeping all the other parameters as in \cite{2009A&A...504..415S}. 
Such a large temperature for the cold gas component is typical of regions of massive star formation \citep[see e.g.][]{2002ApJ...566..931S}.

Using equation (2) of \cite{2009A&A...504..415S}, we additionally derived the H$_2$ column density towards the region, and then used $N(\mathrm{H}_2)/ A_{V} = 0.94 \times 10^{21} \, \mbox{cm}^{-2}\,\mbox{mag}^{-1}$ \citep{1982ApJ...262..590F} to convert it into visual extinction.
At the position of the E1 star we obtain $A_V \approx 4.2$ mag, in perfect agreement with the spectroscopic measurements. In comparison, the ATLASGAL maps provide very high values for the extinction along the lines of sight towards E2, E3 and E4 in the DB10 cluster region, with $A_V \approx 15$ mag.
The extinction values depend on the choice of the dust temperature. For example, choosing $T=30$ K, the estimated $A_{V}$s are almost doubled. Therefore our quoted extinction values from sub-mm emission have to be considered as approximate estimates.
The fact that in DB11 the spectroscopic and sub-mm results are in agreement, while they are not for DB10, might be due to several factors. For example the bright nebula might partially affect our photometric measurements and therefore the reddening estimates. However, a difference of 10 mag in $A_V$ would correspond to an error in the $J-K_{\mathrm{S}}$ color of more than 2 mag, which is too large for these bright sources.
Another possibility is that the differences in the extinction derived for DB10 are related to the beam size of the ATLASGAL maps of $~19\arcsec$. This is comparable in size to the visual radius of the DB10 cluster. Given this comparably large beam, the measurements of the flux at the DB10 position are possibly affected by the emission of the prominent nearby south-west arc and clump. DB11, on the contrary is located in a region of lower sub-mm brightness, with no close prominent features, hence the spectro-photometric results and those from the 870$\umu$m maps are very similar.

\section{Integrated radio continuum flux}
\label{sec:radio}

The radio flux from the two H~\textsc{ii} regions in the CN15/16/17 complex can be used to put additional constraints on the spectral type of the most luminous ionizing source. The amount of ionizing flux is a very steep function of the spectral type \citep{2005A&A...436.1049M}. Consequently, the total ionizing flux within an H~\textsc{ii} region is dominated by the hottest, earliest type star. 

The free electrons inside the H~\textsc{ii} region interact with the ions producing Bremsstrahlung.
This free-free radiation flux, measured at a given wavelength, can be converted into an ionizing flux. Assuming that each energetic photon ionizes an atom (no leakage), that the nebula is optically thin for free-free continuum emission and that ionization and recombination are in equilibrium, the amount of energetic photons and radio continuum emission can be related to each other.

The values of the integrated radio flux at 1.4 GHz are taken from the NVSS catalog \citep{1998AJ....115.1693C}. The ultra compact H~\textsc{ii} region is the brightest at 1.4 GHz, with a flux of 657 mJy, while the flux of the H~\textsc{ii} region associated with DB11 is 572 mJy.
The observed radio flux has first to be converted into an emitted flux by multiplying it with $4\pi d^2$.

We used the relations by \cite{1994ApJS...91..659K} to obtain an estimate of the ionizing fluxes.
Given the estimated distance of $d = 1.2 \pm 0.5 $kpc, (see Sect.~\ref{sec:fsd}) we obtain:
\begin{eqnarray}
\log Q_{0}^{\mathrm{DB11}}[\mathrm{photons}\, \mathrm{s}^{-1}] & = & 46.8_{-0.5}^{+0.3} \nonumber \\
 & & \nonumber \\
\log Q_{0}^{\mathrm{UCH\mathsmaller{II}}}[\mathrm{photons} \,\mathrm{s}^{-1}]  & = & 46.9_{-0.5}^{+0.3} \nonumber
\end{eqnarray}
 where the flux $Q_{0}$ is the number of photons with $\lambda < 912$ \AA~emitted per second.

We compared these values with those predicted by theoretical models of hot stars and derived estimates for the spectral 
types of the ionizing sources inside the DB11 cluster and the Ultra Compact H~\textsc{ii} region. 
Adopting the calculations by \cite{2005A&A...436.1049M}, the expected flux from the latest spectral type available  --an O9.5V--  varies between $\log Q_{0} \sim 47.56$ and $\log Q_{0} \sim 47.88$, depending on the adopted $T_{\mathrm{eff}}$ scale (derived either from models or observations). 
Similarly we considered the ionizing flux computed by \cite{1998ApJ...501..192D} using \cite{1993sssp.book.....K} model atmospheres. These computations extend to lower stellar temperatures than those of \cite{2005A&A...436.1049M}. 
The authors predict an ionizing flux $\log Q_{0} = 47.02$ for a 17.5$M_{\sun}$, solar metallicity, main sequence star, corresponding to $T_{\mathrm{eff}} = 30,000 \,\mathrm{K}$. This flux is consistent with our estimates. No spectral type is quoted by the authors, but from \cite{2005A&A...436.1049M} a $T_{\mathrm{eff}} = 30,000\,\mathrm{K}$ corresponds to an O9.5 class V star, for which, however, their predicted inonizing flux is $\log Q_{0} \sim 47.56$. These differences testify the difficulties in modeling the atmospheres of hot stars. 
Given this uncertainty we cannot constrain the expected spectral type to better than few spectral subclasses, but we can safely assume that our estimated values are consistent with the earliest type of the classified stars being late O or early B.


\section{Nebular emission}
\label{sec:nebem}
The intensity of the diffuse nebular emission can be used to further constrain the spectral type for the brightest stars inside DB10 and DB11. As shown by \cite{2002ApJS..138...35H}, the ratio of the fluxes of the Br$\gamma$ and He~\textsc{i} 2.113$\umu$m lines is an indicator of the temperature of the ionizing source. 

\begin{figure*}
 \centering
  \resizebox{0.9\hsize}{!}{\includegraphics{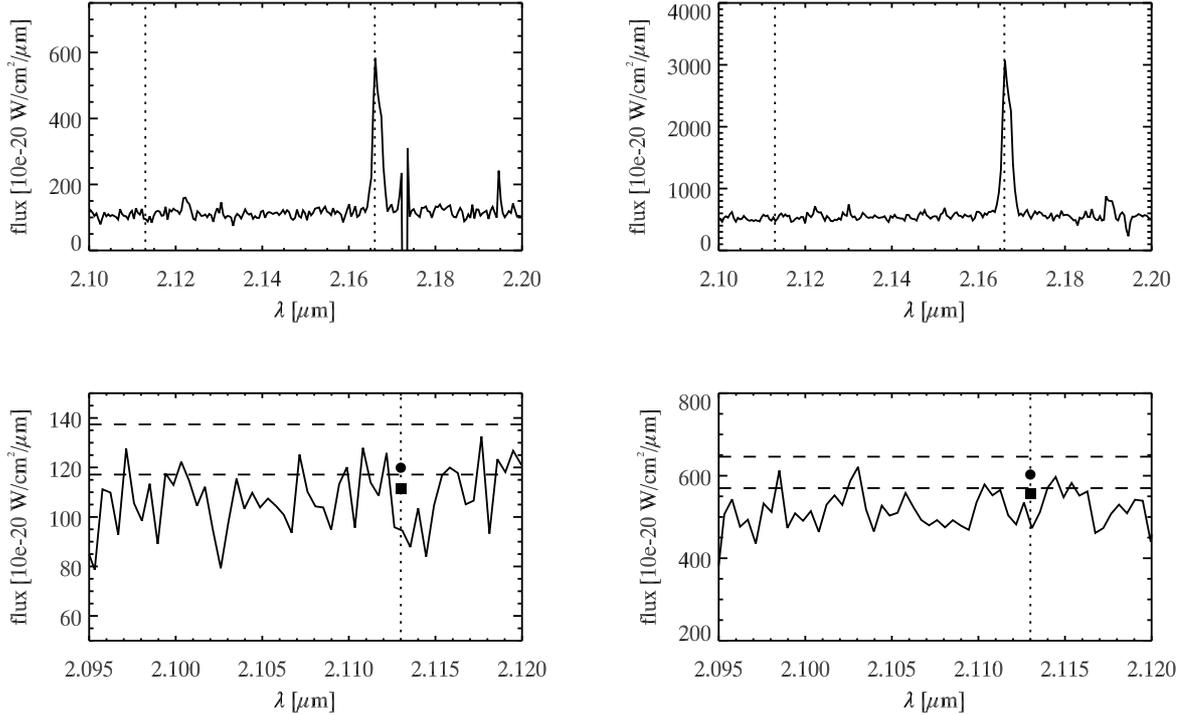}}
\caption{Spectra of the diffuse nebulae in the DB10 (left) and DB11 (right) regions. The positions of the Br$\gamma$ and the He~\textsc{i} 2.113 lines are marked by vertical dotted lines. The lower panels are a zoomed-in version in the wavelength region of the He~\textsc{i} 2.113 line. The horizontal dashed lines represent 1 and 3$\sigma$ levels above the continuum. The squares and the circles represent the expected peak values of the He~\textsc{i} 2.113 line in the case of a 1\% and 3\% He~\textsc{i}/Br$\gamma$ flux ratio, respectively.}
  \label{fig:nebsp}
\end{figure*}

We measured the line fluxes in the nebulae of DB11 and DB10 using the parts of the 2D long-slit spectra which spatially coincide with the nebular regions.
In order to obtain the maximum SNR and thus the best possible constraints on these two fluxes, we combined all five slit positions for which we obtained spectra.
Since exposure times and observing conditions were different among the five slit positions, we first calibrated the 2D frames in flux. We used the spectra of the telluric standard that was observed immediately after or before the given slit. We integrated the measured counts convolved with the 2MASS $K_{\mathrm{S}}$ filter response curve and used the available magnitude for the star to obtain an absolute calibration of the flux for the standard star. By dividing the 2D frames for this calibrated standard spectrum and adjusting for the different exposure times  we obtained 2D flux calibrated frames.
We identified in each 2D frame the regions of diffuse Br$\gamma$ emission associated with either DB10 or DB11, and summed the flux in each region.
The fluxes for the five slits were then co-added to obtain the total nebular flux. The summed spectra for DB10 and DB11 are shown in the top panels of 
Fig.~\ref{fig:nebsp}.


According to \cite{2002ApJS..138...35H} a He~\textsc{i} (2.113)/Br$\gamma$ flux ratio between 1\% and 3\% implies that some late O type stars contribute to the ionization of H~\textsc{ii} regions, and lower ratios imply that no O star is present. In both DB10 and DB11, we find no He~\textsc{i} (2.113) emission, shown in the lower panels of Fig.~\ref{fig:nebsp}. Assuming a similar line shape and FWHM to the Br-gamma line, we plot the expected flux level for the 1\% and 3\% flux ratio cases. We also show the 1$\sigma$ and 3$\sigma$ detection levels, where the standard deviation $\sigma$ of the flux is measured in adjacent featureless parts of the spectrum. In both DB10 and DB11, a 3\% flux ratio can be excluded at the 1-$\sigma$ level, but a 1\% flux ratio cannot be ruled out by our data. We can therefore exclude the presence of early O type stars as the ionizing source in these regions. The earliest spectral type would be early B or late O-type stars.

\section{Mass estimates of the DB10 and DB11 clusters}
\label{sec:db10db11}

\begin{figure*}[htp]
 \centering
  \resizebox{\hsize}{!}{\includegraphics{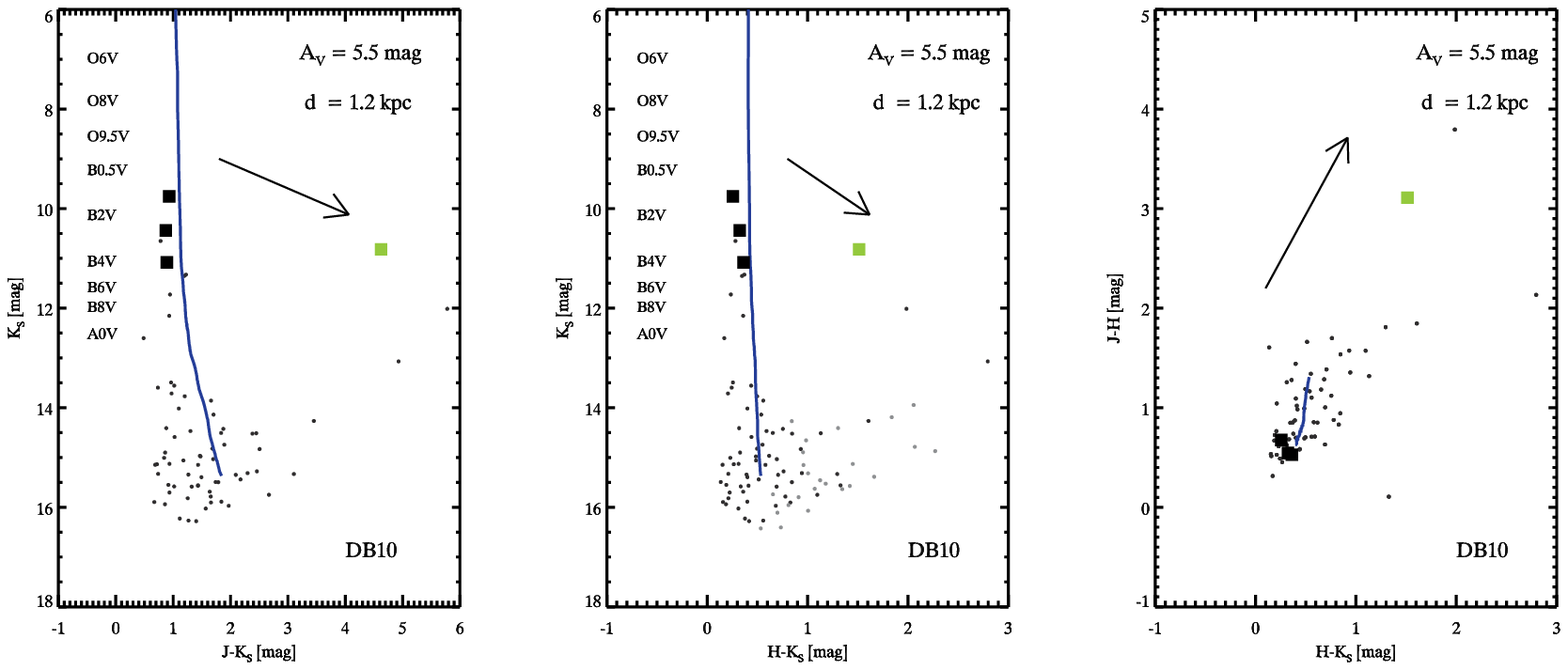}}
  \resizebox{\hsize}{!}{\includegraphics{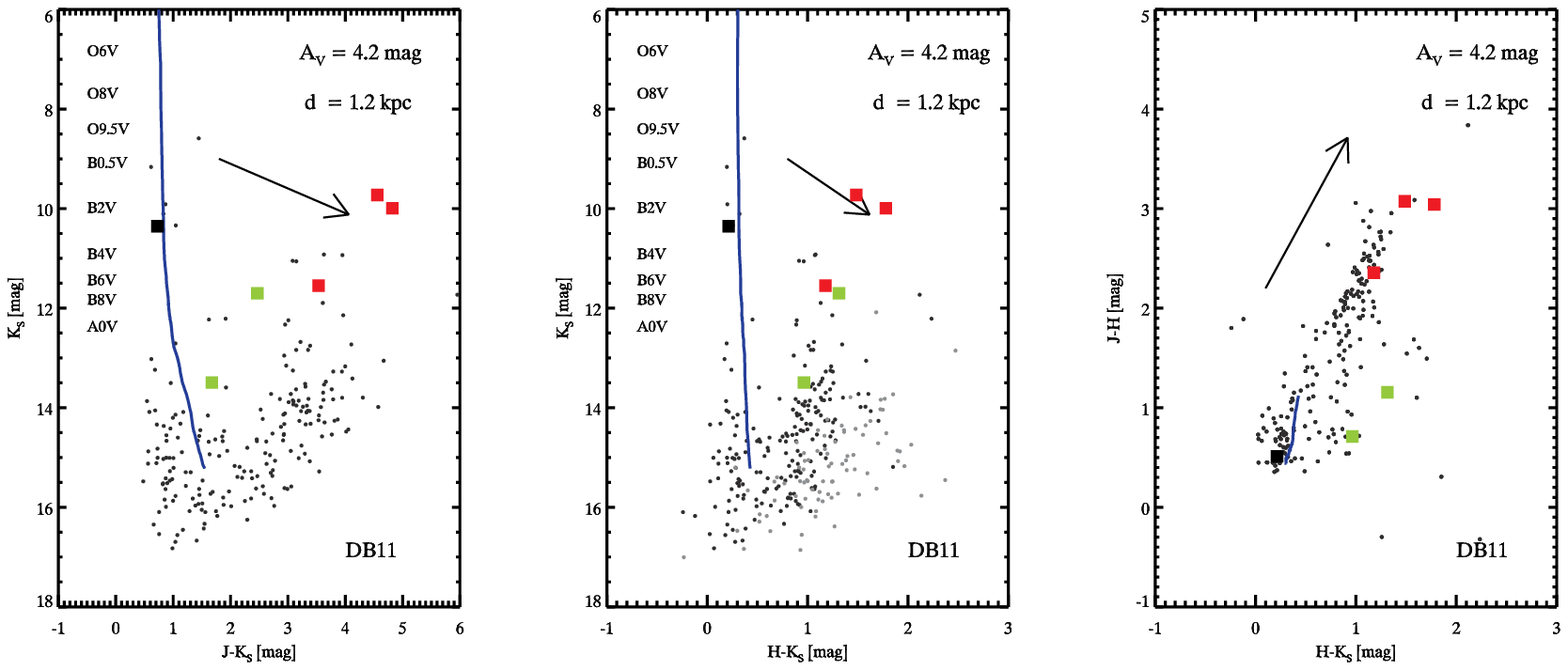}}
 \caption{Color-magnitude and color-color diagrams for DB10 (top) and DB11 (bottom). The blue line is a 1 Myr isochrone from \cite{2001A&A...366..538L}. Spectral types from \cite{2006A&A...457..637M} for O stars and from \cite{1995ApJS..101..117K} for later type stars are shown. Grey dots in the central column panels are stars detected only in $H$ and $K_{\mathrm{S}}$. Squares represent the stars with spectral classification: early types in black, YSOs in green, red giants in red. The arrows indicate the direction of the reddening vector; their lengths correspond to $A_V = 10$ mag.}
 \label{fig:cmd1011}
\end{figure*}

In this section we will focus on the two clusters in the region, DB10 --more embedded and compact-- and DB11 --less embedded, more extended. We consider two circular regions as the clusters areas. These areas correspond to the diffuse nebulosity areas visible in the $K_{\mathrm{S}}$ band image (see Fig.~\ref{fig:Kb_sp}). DB10 is centered at  (RA, Dec)$_{\mathrm{J2000}} = (17^{\mathrm{h}}50^{\mathrm{m}}17.\!^{\mathrm{s}}64, -28\degr53\arcmin41\farcs6)$, $(l,b)_{\mathrm{gal}} = (0.564, -0.854)$, with a visual radius of $21\arcsec$. Object DB11 is centered at (RA, Dec)$_{\mathrm{J2000}} = (17^{\mathrm{h}}50^{\mathrm{m}}23.\!^{\mathrm{s}}79, -28\degr53\arcmin05\farcs1)$, $(l,b)_{\mathrm{gal}} = (0.584, -0.868)$, with a visual radius of $32\arcsec$. At a distance of 1200 pc the radii of the clusters are $\sim 0.12$ pc and $\sim 0.18$ pc for DB10 and DB11 respectively.

The near infrared color-magnitude and color-color diagrams for the two clusters are shown in Fig.~\ref{fig:cmd1011}. Overplotted is a 1 Myr isochrone by \cite{2001A&A...366..538L}.
We have chosen this age based on the presence of the molecular material in the surrounding bubble. This material will be removed by the winds and radiation pressure from massive stars. A slightly older age, up to few Myr might still be possible but we consider 1 Myr as a typical value.
A more precise age determination can be obtained with deeper and higher angular resolution images revealing the pre main sequence population.
We adopt the average value of the spectro-photometric distance obtained from the classified B stars, $d=1.2$ kpc. 
Using the same the spectro-photometric analysis and the \cite{1985ApJ...288..618R} extinction law we derive the extinction towards the clusters. For DB10 we used the average extinction of the stars E2, E3, E4, $A_V = 5.5$ mag ($A_J=1.54$ mag, $A_H=0.96$ mag and $A_{K_{\mathrm{S}}}=0.61$ mag). For DB11 we adopted the value from E1, $A_V = 4.2$ mag ($A_J=1.18$ mag, $A_H=0.74$ mag and $A_{K_{\mathrm{S}}} = 0.47$ mag). 

The positions of the black squares in Fig.~\ref{fig:cmd1011} indicate the magnitudes of the classified objects, with early type stars in black, YSOs in green and red giants in red. 
The bright nebulosity in DB10 hampers the detection of faint $K_\mathrm{S}$ band sources, consequently the diagrams appear quite empty in redder areas.
In DB11 the nebulosity is not as bright and therefore fainter (and redder) sources can be seen in the diagram. 

The classified B stars nicely follow the 1 Myr isochrone. 
The spectroscopically identified YSOs appear redder than the main sequence locus in both the $K_{\mathrm{S}}$ vs $J-K_{\mathrm{S}}$ and the $K_{\mathrm{S}}$ vs $H-K_{\mathrm{S}}$ diagrams. From the color-color diagram, in the case of DB11, it is clear that the redder YSO colors are not due to extinction but are related to near infrared excess from the circumstellar material. In this diagram the two DB11 YSOs (Y2, Y3) are indeed located below the reddening sequence.
The YSO Y2 is identified in the GLIMPSE II catalog as well. With its colors of [5.8]-[8.0] = 1.01 mag and [3.6]-[4.5] = 0.69 mag it is one of the objects identified as a class II YSO in the IRAC color-color diagram of Fig.~\ref{fig:Diagram_ALL}.
In the DB11 diagrams of Fig.~\ref{fig:cmd1011} Y2 is the redder ($J-K_{\mathrm{S}} = 2.47$ mag) of the two marked YSOs.
The YSO in DB10 (Y1) does not show the same near infrared excess. This object might indeed be more embedded as it appears from the high extinction in the color-color diagram. Unfortunately, we could not find a match for this source in the  GLIMPSE II catalog, but the very red and featureless spectrum indicates a class I type.
The three red giants classified in the DB11 area (RG1, RG2, RG4) all appear to be very reddened background sources, most likely belonging to the Galactic Bulge.

The cluster DB11 was already studied in \cite{2003A&A...408..127D} without the help of spectroscopy and with very different results.
After performing statistical field subtraction, the authors identified a much redder main sequence for the region with a color of $H-K_\mathrm{S} \sim 1.12$ mag. Fitting this sequence they derived an $A_V$ of 15 mag ($A_{K_{\mathrm{S}}} = 1.68$ mag). 
The classified B star in DB11 (E1) shows, in contrast, a much bluer $H-K_\mathrm{S}$ color of only 0.21 mag and a much lower extinction $A_V = 4.6$ mag ($A_{K_{\mathrm{S}}} = 0.47$).
Three of the classified objects lie in the region $H-K_\mathrm{S} \sim 1.12$. One of them is a red giant and the other two are the Y2 and Y3 YSOs. The presence of the first is an indication that this part of the DB11 color magnitude diagram is contaminated by red giants. Moreover the color-color diagram shows that the two YSOs clearly have a $K_\mathrm{S}$ band excess, being displaced from the main sequence by $\sim 0.6$ mag. If the main sequence of DB11 in the $K_\mathrm{S}$ vs $H-K_\mathrm{S}$ diagram were indeed at $H-K_\mathrm{S} \sim 1.12$ mag we would expect the two YSOs to be $\sim 0.6$ mag redder than they are.

Assuming that the 10th brightest star of the group that they identified as cluster members was a B0V star, \cite{2003A&A...408..127D} derived a distance to the cluster of 7.6 kpc.
This scenario was corroborated by the value of the kinematic distance for the associated H~\textsc{ii} region. \cite{1997ApJ...488..224K} estimated a velocity along the line of sight of 15 km/s. This value of the velocity can be translated into a distance once a model of the Galactic Rotation curve is adopted. Given the line of sight for this region --very close to the Galactic Center-- a very small radial velocity is expected for all orbits but those very close to the Galactic center. The derived kinematic distances are indeed 7.5 kpc (near) and 9.5 kpc (far) using the Galactic rotation model by \cite{1993A&A...275...67B} with a distance of the Sun from the Galactic Center $R_{0} = 8.5$ kpc and an orbital velocity at $R_{0}$, $\theta = 220$ km s$^{-1}$. The kinematic distance value is very uncertain though. For objects that are moving almost perpendicular to the line of sight, the peculiar orbital motions and the internal motion of the H~\textsc{ii} region can dominate the value of the line of sight velocity.

Our spectroscopic observations give contrasting results for the distance of the cluster. We argue that the observed main sequence of \cite{2003A&A...408..127D} might be a result of the subtraction technique, related to the difficulties of choosing a representative control field, when looking towards the Galactic Center. Along such lines of sight the extinction pattern might be very patchy and strong differences can be observed between adjacent areas. This is also the case for the CN15/16/17 region where, as visible in Fig.~\ref{fig:YSOreg}, local concentration of cold dust correspond to large values of extinction. The area centered on DB11 has been cleared of the molecular material by the early type stars with their winds. This material accumulates in the rims of the bubble, but not directly at the cluster position; as a consequence extinction is smaller in the cluster's area. Therefore it might be possible that a larger fraction of background stars is visible along the DB11 line of sight, when compared to the neighboring regions used by \cite{2003A&A...408..127D} for statistical field subtraction (see also Fig.~\ref{fig:imreg}). This would result in an under-subtraction of background giants in the DB11 color-magnitude diagram.

If the 10-th most massive star of the cluster would have been of B0V type --as proposed by \cite{2003A&A...408..127D}-- the presence of several O stars would be expected. We ruled this out by showing that both the nebular emission and the radio flux are consistent with the brightest stars being B-type.
In our target selection for spectroscopy we have chosen the brightest stars in the region and identified them as early B stars. Nevertheless we might have missed some earlier-type stars, because they could be more extincted and appear fainter. However, in this case the diffuse emission and the radio data would not have been consistent with DB11 hosting at most early B or very late O stars and being at a distance of 1.2 kpc.
A distance of 7.5 kpc would imply the presence of early O stars, which would have been detected through a much larger He~\textsc{i} nebular emission at 2.113$\umu$m (see Sect.~\ref{sec:nebem}).

Assuming that we identified the most massive star in DB10 (B1V, $\sim 12 M_{\sun}$) and DB11 (B0V, $\sim 15 M_{\sun}$) and using the relation between the mass of the most massive star in a cluster and its total mass by \cite{2003ASPC..287...65L}, we can give an estimate of the clusters masses. We obtain $M_{\mathrm{DB10}} \approx 170 M_{\sun}$ and $M_{\mathrm{DB11}} \approx 275 M_{\sun}$, both comparable to the mass of the $\sigma$ Ori cluster in the Orion OB1b association, $M_{\sigma\,\mathrm{Ori}} \approx 225 \pm 30 M_{\sun}$ \citep{2004AJ....128.2316S}.


\section{Discussion and Conclusions}
\label{sec:conc}

We performed a multiple wavelength study of the young star forming region CN15/16/17 towards the Galactic Center.
We derived a consistent picture of the region combining near infrared spectroscopy, near and mid infrared photometry, sub-mm continuum emission and integrated radio fluxes.

The region hosts two near infrared clusters (DB10 and DB11, visible in the SofI images) and a loose association of YSOs identified in the GLIMPSE~II point-sources catalog.
The DB11 cluster is associated with an H~\textsc{ii} region visible in the NVSS radio continuum maps. Another H~\textsc{ii} region is also visible, associated to a small group of very embedded mid infrared bright sources.
From ATLASGAL sub-mm contours we identify an arc-like structure in the south-western part of the region. This arc of cold, dense material is associated with the southern group of YSOs.

Considering this collection of evidence from multiple wavelength regimes, a qualitative picture of the star formation history of the region can be traced.
The central cluster, DB11, is the oldest star formation site, followed by the younger cluster DB10 and the youngest Ultra Compact H~\textsc{ii} region. Star formation is proceeding sequentially from DB11 outwards. This is corroborated by the presence of the cold dust arc in the south, which is associated, in turn, with the most populous group of YSOs of the region. In addition, two small sub-groups of YSOs are visible in the north and north-west, again associated with cold dust emission. These two smaller concentrations are at a similar projected distance from DB11 as the south-west arc. 
However, the age determination of the clusters is quite uncertain and, e.g. spectroscopic identification of pre-main sequence members is required to place DB11 and DB10 on a more robust age scale. 
Further investigation is required to understand whether the apparently younger events have been triggered by the older ones \citep[see e.g][as examples of analysis of triggered star formation in mid infrared bubbles]{2010ApJ...716.1478W, 2010A&A...523A...6D}, or star formation has been seeded independently in multiple sites of the parental molecular cloud, in analogy, e.g., to what is observed in the Rosette Molecular Complex \citep{2005ApJ...620..816L,2008ApJ...672..861R}.

Through spectroscopic characterization of the brightest members of the DB10 and DB11 clusters, we obtained a spectro-photometric distance to the region as well as the extinction value towards the clusters.
We found that the region is at $d=1.2\pm0.5$ kpc, i.e. much closer than previously thought when only $H$ and $K_\mathrm{S}$ photometry was used for the analysis of the DB11 cluster.
From the mass of the most massive stars we estimate the stellar cluster masses to be $M_{\mathrm{DB10}} \approx 170 M_{\sun}$ and $M_{\mathrm{DB11}} \approx 275 M_{\sun}$.

This work affirms the importance of the multi-wavelength approach to studies of young star forming regions. We illustrate that the combination of imaging and spectroscopy is essential to the understanding of embedded stellar populations.


\begin{acknowledgements}
  M.G. is grateful to Sarah Ragan, Katharine Johnston and Rory Holmes (MPIA) for helping with the manuscript. 
  A.S. acknowledges generous support from the DFG Emmy Noether programme, grant STO 496/3-1.
  D.G. kindly acknowledges the German Aerospace Center (DLR) and the German Federal Ministry for Economics and Technology (BMWi) for their support through grant 50 OR 0908.
\end{acknowledgements}

\bibliographystyle{aa} 
\bibliography{biblioCN15_P} 

\begin{thebibliography}{51}
\expandafter\ifx\csname natexlab\endcsname\relax\def\natexlab#1{#1}\fi

\bibitem[{{Allen} {et~al.}(2004){Allen}, {Calvet}, {D'Alessio}, {Merin},
  {Hartmann}, {Megeath}, {Gutermuth}, {Muzerolle}, {Pipher}, {Myers}, \&
  {Fazio}}]{2004ApJS..154..363A}
{Allen}, L.~E., {Calvet}, N., {D'Alessio}, P., {et~al.} 2004, \apjs, 154, 363

\bibitem[{{Arnaboldi} {et~al.}(2007){Arnaboldi}, {Neeser}, {Parker}, {Rosati},
  {Lombardi}, {Dietrich}, \& {Hummel}}]{2007Msngr.127...28A}
{Arnaboldi}, M., {Neeser}, M.~J., {Parker}, L.~C., {et~al.} 2007, The
  Messenger, 127, 28

\bibitem[{{Benjamin} {et~al.}(2003){Benjamin}, {Churchwell}, {Babler}, {Bania},
  {Clemens}, {Cohen}, {Dickey}, {Indebetouw}, {Jackson}, {Kobulnicky},
  {Lazarian}, {Marston}, {Mathis}, {Meade}, {Seager}, {Stolovy}, {Watson},
  {Whitney}, {Wolff}, \& {Wolfire}}]{2003PASP..115..953B}
{Benjamin}, R.~A., {Churchwell}, E., {Babler}, B.~L., {et~al.} 2003, \pasp,
  115, 953

\bibitem[{{Bik} {et~al.}(2012){Bik}, {Henning}, {Stolte}, {Brandner},
  {Gouliermis}, {Gennaro}, {Pasquali}, {Rochau}, {Beuther}, {Ageorges},
  {Seifert}, {Wang}, \& {Kudryavtseva}}]{2012ApJ...744...87B}
{Bik}, A., {Henning}, T., {Stolte}, A., {et~al.} 2012, \apj, 744, 87

\bibitem[{{Bik} {et~al.}(2005){Bik}, {Kaper}, {Hanson}, \&
  {Smits}}]{2005A&A...440..121B}
{Bik}, A., {Kaper}, L., {Hanson}, M.~M., \& {Smits}, M. 2005, \aap, 440, 121

\bibitem[{{Brand} \& {Blitz}(1993)}]{1993A&A...275...67B}
{Brand}, J. \& {Blitz}, L. 1993, \aap, 275, 67

\bibitem[{{Churchwell} {et~al.}(2009){Churchwell}, {Babler}, {Meade},
  {Whitney}, {Benjamin}, {Indebetouw}, {Cyganowski}, {Robitaille}, {Povich},
  {Watson}, \& {Bracker}}]{2009PASP..121..213C}
{Churchwell}, E., {Babler}, B.~L., {Meade}, M.~R., {et~al.} 2009, \pasp, 121,
  213

\bibitem[{{Churchwell} {et~al.}(2006){Churchwell}, {Povich}, {Allen}, {Taylor},
  {Meade}, {Babler}, {Indebetouw}, {Watson}, {Whitney}, {Wolfire}, {Bania},
  {Benjamin}, {Clemens}, {Cohen}, {Cyganowski}, {Jackson}, {Kobulnicky},
  {Mathis}, {Mercer}, {Stolovy}, {Uzpen}, {Watson}, \&
  {Wolff}}]{2006ApJ...649..759C}
{Churchwell}, E., {Povich}, M.~S., {Allen}, D., {et~al.} 2006, \apj, 649, 759

\bibitem[{{Churchwell} {et~al.}(2007){Churchwell}, {Watson}, {Povich},
  {Taylor}, {Babler}, {Meade}, {Benjamin}, {Indebetouw}, \&
  {Whitney}}]{2007ApJ...670..428C}
{Churchwell}, E., {Watson}, D.~F., {Povich}, M.~S., {et~al.} 2007, \apj, 670,
  428

\bibitem[{{Clark} \& {Steele}(2000)}]{2000A&AS..141...65C}
{Clark}, J.~S. \& {Steele}, I.~A. 2000, \aaps, 141, 65

\bibitem[{{Condon} {et~al.}(1998){Condon}, {Cotton}, {Greisen}, {Yin},
  {Perley}, {Taylor}, \& {Broderick}}]{1998AJ....115.1693C}
{Condon}, J.~J., {Cotton}, W.~D., {Greisen}, E.~W., {et~al.} 1998, \aj, 115,
  1693

\bibitem[{{Cooley} \& {Tukey}(1965)}]{FFTart}
{Cooley}, J.~W. \& {Tukey}, J.~W. 1965, Math. Comp., 19, 297

\bibitem[{{Deharveng} {et~al.}(2010){Deharveng}, {Schuller}, {Anderson},
  {Zavagno}, {Wyrowski}, {Menten}, {Bronfman}, {Testi}, {Walmsley}, \&
  {Wienen}}]{2010A&A...523A...6D}
{Deharveng}, L., {Schuller}, F., {Anderson}, L.~D., {et~al.} 2010, \aap, 523,
  A6

\bibitem[{{Devillard}(2001)}]{Devillard:2001fk}
{Devillard}, N. 2001, in Astronomical Society of the Pacific Conference Series,
  Vol. 238, Astronomical Data Analysis Software and Systems X,, ed.
  {F.~R.~Harnden Jr., F.~A.~Primini, \& H.~E.~Payne}, 525

\bibitem[{{Diaz-Miller} {et~al.}(1998){Diaz-Miller}, {Franco}, \&
  {Shore}}]{1998ApJ...501..192D}
{Diaz-Miller}, R.~I., {Franco}, J., \& {Shore}, S.~N. 1998, \apj, 501, 192

\bibitem[{{Dutra} \& {Bica}(2000)}]{2000A&A...359L...9D}
{Dutra}, C.~M. \& {Bica}, E. 2000, \aap, 359, L9

\bibitem[{{Dutra} {et~al.}(2003){Dutra}, {Ortolani}, {Bica}, {Barbuy},
  {Zoccali}, \& {Momany}}]{2003A&A...408..127D}
{Dutra}, C.~M., {Ortolani}, S., {Bica}, E., {et~al.} 2003, \aap, 408, 127

\bibitem[{{Epchtein} {et~al.}(1999){Epchtein}, {Deul}, {Derriere},
  {Borsenberger}, {Egret}, {Simon}, {Alard}, {Bal{\'a}zs}, {de Batz}, {Cioni},
  {Copet}, {Dennefeld}, {Forveille}, {Fouqu{\'e}}, {Garz{\'o}n}, {Habing},
  {Holl}, {Hron}, {Kimeswenger}, {Lacombe}, {Le Bertre}, {Loup}, {Mamon},
  {Omont}, {Paturel}, {Persi}, {Robin}, {Rouan}, {Tiph{\`e}ne}, {Vauglin}, \&
  {Wagner}}]{1999A&A...349..236E}
{Epchtein}, N., {Deul}, E., {Derriere}, S., {et~al.} 1999, \aap, 349, 236

\bibitem[{{Frerking} {et~al.}(1982){Frerking}, {Langer}, \&
  {Wilson}}]{1982ApJ...262..590F}
{Frerking}, M.~A., {Langer}, W.~D., \& {Wilson}, R.~W. 1982, \apj, 262, 590

\bibitem[{{Hanson} {et~al.}(2005){Hanson}, {Kudritzki}, {Kenworthy}, {Puls}, \&
  {Tokunaga}}]{2005ApJS..161..154H}
{Hanson}, M.~M., {Kudritzki}, R.-P., {Kenworthy}, M.~A., {Puls}, J., \&
  {Tokunaga}, A.~T. 2005, \apjs, 161, 154

\bibitem[{{Hanson} {et~al.}(2002){Hanson}, {Luhman}, \&
  {Rieke}}]{2002ApJS..138...35H}
{Hanson}, M.~M., {Luhman}, K.~L., \& {Rieke}, G.~H. 2002, \apjs, 138, 35

\bibitem[{{Horne}(1986)}]{1986PASP...98..609H}
{Horne}, K. 1986, \pasp, 98, 609

\bibitem[{{Izzo} {et~al.}(2004){Izzo}, {Kornweibel}, {McKay}, {Palsa}, {Peron},
  \& {Taylor}}]{2004Msngr.117...33I}
{Izzo}, C., {Kornweibel}, N., {McKay}, D., {et~al.} 2004, The Messenger, 117,
  33

\bibitem[{{Kenyon} \& {Hartmann}(1995)}]{1995ApJS..101..117K}
{Kenyon}, S.~J. \& {Hartmann}, L. 1995, \apjs, 101, 117

\bibitem[{{Kuchar} \& {Clark}(1997)}]{1997ApJ...488..224K}
{Kuchar}, T.~A. \& {Clark}, F.~O. 1997, \apj, 488, 224

\bibitem[{{Kurtz} {et~al.}(1994){Kurtz}, {Churchwell}, \&
  {Wood}}]{1994ApJS...91..659K}
{Kurtz}, S., {Churchwell}, E., \& {Wood}, D.~O.~S. 1994, \apjs, 91, 659

\bibitem[{{Kurucz}(1993)}]{1993sssp.book.....K}
{Kurucz}, R.~L. 1993, {SYNTHE spectrum synthesis programs and line data}, ed.
  {Kurucz, R.~L.}

\bibitem[{{Larson}(2003)}]{2003ASPC..287...65L}
{Larson}, R.~B. 2003, in Astronomical Society of the Pacific Conference Series,
  Vol. 287, Galactic Star Formation Across the Stellar Mass Spectrum, ed.
  {J.~M.~De Buizer \& N.~S.~van der Bliek}, 65--80

\bibitem[{{Lejeune} \& {Schaerer}(2001)}]{2001A&A...366..538L}
{Lejeune}, T. \& {Schaerer}, D. 2001, \aap, 366, 538

\bibitem[{{Li} \& {Smith}(2005)}]{2005ApJ...620..816L}
{Li}, J.~Z. \& {Smith}, M.~D. 2005, \apj, 620, 816

\bibitem[{{Longmore} {et~al.}(2012){Longmore}, {Rathborne}, {Bastian}, {Alves},
  {Ascenso}, {Bally}, {Testi}, {Longmore}, {Battersby}, {Bressert}, {Purcell},
  {Walsh}, {Jackson}, {Foster}, {Molinari}, {Meingast}, {Amorim}, {Lima},
  {Marques}, {Moitinho}, {Pinhao}, {Rebordao}, \&
  {Santos}}]{2012ApJ...746..117L}
{Longmore}, S.~N., {Rathborne}, J., {Bastian}, N., {et~al.} 2012, \apj, 746,
  117

\bibitem[{{Lucas} {et~al.}(2008){Lucas}, {Hoare}, {Longmore}, {Schr{\"o}der},
  {Davis}, {Adamson}, {Bandyopadhyay}, {de Grijs}, {Smith}, {Gosling},
  {Mitchison}, {G{\'a}sp{\'a}r}, {Coe}, {Tamura}, {Parker}, {Irwin}, {Hambly},
  {Bryant}, {Collins}, {Cross}, {Evans}, {Gonzalez-Solares}, {Hodgkin},
  {Lewis}, {Read}, {Riello}, {Sutorius}, {Lawrence}, {Drew}, {Dye}, \&
  {Thompson}}]{2008MNRAS.391..136L}
{Lucas}, P.~W., {Hoare}, M.~G., {Longmore}, A., {et~al.} 2008, \mnras, 391, 136

\bibitem[{{Marengo} {et~al.}(2008){Marengo}, {Reiter}, \&
  {Fazio}}]{2008AIPC.1001..331M}
{Marengo}, M., {Reiter}, M., \& {Fazio}, G.~G. 2008, in American Institute of
  Physics Conference Series, Vol. 1001, Evolution and Nucleosynthesis in AGB
  Stars, ed. {R.~Guandalini, S.~Palmerini, \& M.~Busso}, 331--338

\bibitem[{{Martins} \& {Plez}(2006)}]{2006A&A...457..637M}
{Martins}, F. \& {Plez}, B. 2006, \aap, 457, 637

\bibitem[{{Martins} {et~al.}(2005){Martins}, {Schaerer}, \&
  {Hillier}}]{2005A&A...436.1049M}
{Martins}, F., {Schaerer}, D., \& {Hillier}, D.~J. 2005, \aap, 436, 1049

\bibitem[{{Mathis}(1990)}]{1990ARA&A..28...37M}
{Mathis}, J.~S. 1990, \araa, 28, 37

\bibitem[{{Megeath} {et~al.}(2004){Megeath}, {Allen}, {Gutermuth}, {Pipher},
  {Myers}, {Calvet}, {Hartmann}, {Muzerolle}, \& {Fazio}}]{2004ApJS..154..367M}
{Megeath}, S.~T., {Allen}, L.~E., {Gutermuth}, R.~A., {et~al.} 2004, \apjs,
  154, 367

\bibitem[{{Minniti} {et~al.}(2010){Minniti}, {Lucas}, {Emerson}, {Saito},
  {Hempel}, {Pietrukowicz}, {Ahumada}, {Alonso}, {Alonso-Garcia}, {Arias},
  {Bandyopadhyay}, {Barb{\'a}}, {Barbuy}, {Bedin}, {Bica}, {Borissova},
  {Bronfman}, {Carraro}, {Catelan}, {Clari{\'a}}, {Cross}, {de Grijs},
  {D{\'e}k{\'a}ny}, {Drew}, {Fari{\~n}a}, {Feinstein}, {Fern{\'a}ndez
  Laj{\'u}s}, {Gamen}, {Geisler}, {Gieren}, {Goldman}, {Gonzalez}, {Gunthardt},
  {Gurovich}, {Hambly}, {Irwin}, {Ivanov}, {Jord{\'a}n}, {Kerins}, {Kinemuchi},
  {Kurtev}, {L{\'o}pez-Corredoira}, {Maccarone}, {Masetti}, {Merlo},
  {Messineo}, {Mirabel}, {Monaco}, {Morelli}, {Padilla}, {Palma}, {Parisi},
  {Pignata}, {Rejkuba}, {Roman-Lopes}, {Sale}, {Schreiber}, {Schr{\"o}der},
  {Smith}, {Sodr{\'e}}, {Soto}, {Tamura}, {Tappert}, {Thompson}, {Toledo},
  {Zoccali}, \& {Pietrzynski}}]{2010NewA...15..433M}
{Minniti}, D., {Lucas}, P.~W., {Emerson}, J.~P., {et~al.} 2010, \na, 15, 433

\bibitem[{{Morris} \& {Serabyn}(1996)}]{1996ARA&A..34..645M}
{Morris}, M. \& {Serabyn}, E. 1996, \araa, 34, 645

\bibitem[{{Portegies Zwart} {et~al.}(2001){Portegies Zwart}, {Makino},
  {McMillan}, \& {Hut}}]{2001ApJ...546L.101P}
{Portegies Zwart}, S.~F., {Makino}, J., {McMillan}, S.~L.~W., \& {Hut}, P.
  2001, \apjl, 546, L101

\bibitem[{{Rayner} {et~al.}(2009){Rayner}, {Cushing}, \&
  {Vacca}}]{2009ApJS..185..289R}
{Rayner}, J.~T., {Cushing}, M.~C., \& {Vacca}, W.~D. 2009, \apjs, 185, 289

\bibitem[{{Rieke} \& {Lebofsky}(1985)}]{1985ApJ...288..618R}
{Rieke}, G.~H. \& {Lebofsky}, M.~J. 1985, \apj, 288, 618

\bibitem[{{Rom{\'a}n-Z{\'u}{\~n}iga} {et~al.}(2008){Rom{\'a}n-Z{\'u}{\~n}iga},
  {Elston}, {Ferreira}, \& {Lada}}]{2008ApJ...672..861R}
{Rom{\'a}n-Z{\'u}{\~n}iga}, C.~G., {Elston}, R., {Ferreira}, B., \& {Lada},
  E.~A. 2008, \apj, 672, 861

\bibitem[{{Schuller} {et~al.}(2009){Schuller}, {Menten}, {Contreras},
  {Wyrowski}, {Schilke}, {Bronfman}, {Henning}, {Walmsley}, {Beuther},
  {Bontemps}, {Cesaroni}, {Deharveng}, {Garay}, {Herpin}, {Lefloch}, {Linz},
  {Mardones}, {Minier}, {Molinari}, {Motte}, {Nyman}, {Reveret}, {Risacher},
  {Russeil}, {Schneider}, {Testi}, {Troost}, {Vasyunina}, {Wienen}, {Zavagno},
  {Kovacs}, {Kreysa}, {Siringo}, \& {Wei{\ss}}}]{2009A&A...504..415S}
{Schuller}, F., {Menten}, K.~M., {Contreras}, Y., {et~al.} 2009, \aap, 504, 415

\bibitem[{{Sherry} {et~al.}(2004){Sherry}, {Walter}, \&
  {Wolk}}]{2004AJ....128.2316S}
{Sherry}, W.~H., {Walter}, F.~M., \& {Wolk}, S.~J. 2004, \aj, 128, 2316

\bibitem[{{Skrutskie} {et~al.}(2006){Skrutskie}, {Cutri}, {Stiening},
  {Weinberg}, {Schneider}, {Carpenter}, {Beichman}, {Capps}, {Chester},
  {Elias}, {Huchra}, {Liebert}, {Lonsdale}, {Monet}, {Price}, {Seitzer},
  {Jarrett}, {Kirkpatrick}, {Gizis}, {Howard}, {Evans}, {Fowler}, {Fullmer},
  {Hurt}, {Light}, {Kopan}, {Marsh}, {McCallon}, {Tam}, {Van Dyk}, \&
  {Wheelock}}]{Skrutskie:2006uq}
{Skrutskie}, M.~F., {Cutri}, R.~M., {Stiening}, R., {et~al.} 2006, \aj, 131,
  1163

\bibitem[{{Sridharan} {et~al.}(2002){Sridharan}, {Beuther}, {Schilke},
  {Menten}, \& {Wyrowski}}]{2002ApJ...566..931S}
{Sridharan}, T.~K., {Beuther}, H., {Schilke}, P., {Menten}, K.~M., \&
  {Wyrowski}, F. 2002, \apj, 566, 931

\bibitem[{{Stetson}(1987)}]{Stetson:1987qy}
{Stetson}, P.~B. 1987, \pasp, 99, 191

\bibitem[{{Walsh} {et~al.}(1998){Walsh}, {Burton}, {Hyland}, \&
  {Robinson}}]{1998MNRAS.301..640W}
{Walsh}, A.~J., {Burton}, M.~G., {Hyland}, A.~R., \& {Robinson}, G. 1998,
  \mnras, 301, 640

\bibitem[{{Watson} {et~al.}(2010){Watson}, {Hanspal}, \&
  {Mengistu}}]{2010ApJ...716.1478W}
{Watson}, C., {Hanspal}, U., \& {Mengistu}, A. 2010, \apj, 716, 1478

\bibitem[{{Yusef-Zadeh} {et~al.}(2009){Yusef-Zadeh}, {Hewitt}, {Arendt},
  {Whitney}, {Rieke}, {Wardle}, {Hinz}, {Stolovy}, {Lang}, {Burton}, \&
  {Ramirez}}]{2009ApJ...702..178Y}
{Yusef-Zadeh}, F., {Hewitt}, J.~W., {Arendt}, R.~G., {et~al.} 2009, \apj, 702,
  178

\end{thebibliography}

\end{document}